\documentclass[graybox, nosecnum]{svmult}

\usepackage{mathptmx}       
\usepackage{helvet}         
\usepackage{courier}        
\usepackage{type1cm}        
\usepackage{makeidx}         
\usepackage{graphicx}        
\usepackage{multicol}        
\usepackage[bottom]{footmisc}
\usepackage{hyperref}        
\usepackage{soul}            
\hypersetup{colorlinks=true,urlcolor=blue}
\usepackage[square,numbers]{natbib}
\makeindex             
\usepackage{amsmath}
\usepackage{amssymb}
\usepackage{epsfig}
\usepackage{bm}

\begin{document}
\title*{Testing the nature of dark compact objects with gravitational waves}
\author{Elisa Maggio, Paolo Pani, Guilherme Raposo}
\institute{Elisa Maggio \at Dipartimento di Fisica, ``Sapienza" Universit\`a di Roma \& Sezione INFN Roma1, Piazzale Aldo Moro 5, \email{elisa.maggio@uniroma1.it}
\and Paolo Pani \at Dipartimento di Fisica, ``Sapienza" Universit\`a di Roma \& Sezione INFN Roma1, Piazzale Aldo Moro 5, \email{paolo.pani@uniroma1.it}
\and
Guilherme Raposo\at Dipartimento di Fisica, ``Sapienza" Universit\`a di Roma \& Sezione INFN Roma1, Piazzale Aldo Moro 5, \email{guilherme.raposo@uniroma1.it}}
%
%
\maketitle
\abstract{
Within Einstein's theory of gravity, any compact object heavier than a few solar masses must be a black hole. Any observation showing otherwise would imply either new physics beyond General Relativity or new exotic matter fields beyond the Standard Model, and might provide a portal to understand some puzzling properties of a black hole. 
We give a short overview on tests of the nature of dark compact objects with present and future gravitational-wave observations, including inspiral tests of the multipolar structure of compact objects and of their tidal deformability, ringdown tests, and searches for near-horizon structures with gravitational-wave echoes.
}
\section{Keywords} 
gravitational waves; black holes; tests of gravity;

\newpage
\section{Introduction}
Originally considered just as bizarre solutions to Einstein's theory of General Relativity~(GR), black holes~(BHs) have now acquired a prominent role in astrophysics, theoretical physics, and high-energy physics at large, to a level in which ``BH physics" is living a new golden age. Gravitational-wave~(GW) observations have provided the strongest and most direct evidence to date for the existence of BHs with masses ranging from few to hundred solar masses. Meanwhile, progress in modelling the electromagnetic emission from accreting BHs and new X-ray and VLBI facilities have provided new probes of the region near dark compact objects, both in the mass range explored by LIGO and Virgo and in the supermassive range, i.e. million to billion solar masses.
All observations so far are beautifully compatible with the predictions of GR and with the so-called ``Kerr hypothesis", namely that {\it any} compact object with mass larger than a few solar masses is described by the Kerr BH metric, as predicted by some remarkable uniqueness and ``no-hair" theorems established within GR.

Given this state of affairs and the robustness (both at astrophysical and observational level) of the BH picture, it is natural to question the motivation for further tests of the nature of compact objects.
In fact, in synergy with GW-based tests of gravity discussed in other chapters, tests of the nature of compact objects are emerging as one of the cornerstones of strong-gravity research, for several theoretical and phenomenological reasons:

\begin{itemize}
    \item One of the most urgent open problems in theoretical physics is the so-called information-loss paradox~\cite{Mathur:2009hf}, which is related to loss of unitary at the end of the BH evaporation due to Hawking's radiation. This problem should be hopefully resolved within a consistent quantum gravity theory. Several attempts to address this issue (most notably the fuzzball scenario in string theory~\cite{Lunin:2002qf,Mazur:2004fk, Mathur:2005zp, Mathur:2008nj,Mathur:2009hf}, but also nonlocal effects~\cite{Giddings:1992hh,Giddings:2009ae,Giddings:2012bm,
Giddings:2016tla,Giddings:2017mym}, firewalls~\cite{Almheiri:2012rt}, etc.) predict drastic changes at the horizon relative to the classical BH picture, regardless of the curvature of the object.
    \item Although within classical GR nothing special or dramatic happens in the vicinity of the BH horizon, the BH interior is pathological. It harbors curvature singularities (where Einstein's theory breaks down), and may contain closed-timelike curves (which violate causality), and Cauchy horizons. Some of the attempts to regularize the BH interior also affect the near-horizon structure and would leave some imprint in the exterior.
    \item From a more phenomenological standpoint, BHs and neutron stars might be just two ``species" of a larger zoo of compact objects. New species might have very different properties that can be used to devise precision searches with current and future experiments. In this context it is interesting that current LIGO/Virgo observations (especially the recent GW190814~\cite{Abbott:2020khf} and GW190521~\cite{Abbott:2020tfl,Abbott:2020mjq}, respectively in the lower-mass and upper-mass gap forbidden for standard stellar-origin BHs) do not exclude the possibility that some of the GW mergers involve exotic objects~\cite{CalderonBustillo:2020srq}.
    \item Under general conditions, Penrose's  theorem~\cite{Penrose:1964wq} implies that an apparent horizon always hides a curvature singularity. Thus, BHs are not only a unique prediction of Einstein's theory, they are in fact crucial for the self-consistency of the latter. On general grounds, the evidence supporting the existence of horizons should be properly {\it quantified} in the most accurate way. This implies, on the one hand, devising model-agnostic tests of the ``BH-ness" of compact astrophysical sources and, on the other hand, confronting the BH scenarios with more exotic ones, for example using Bayesian model selection.
\end{itemize}

This chapter is devoted to an overview of tests of dark compact objects using GW probes, a topic that has acquired significant attention in the last few years. We refer to other chapters of the book for electromagnetic tests, which are complementary to GW ones in several ways.

\section{Models of exotic compact objects}
\begin{figure}[t]
\centering
\includegraphics[width=0.85\textwidth]{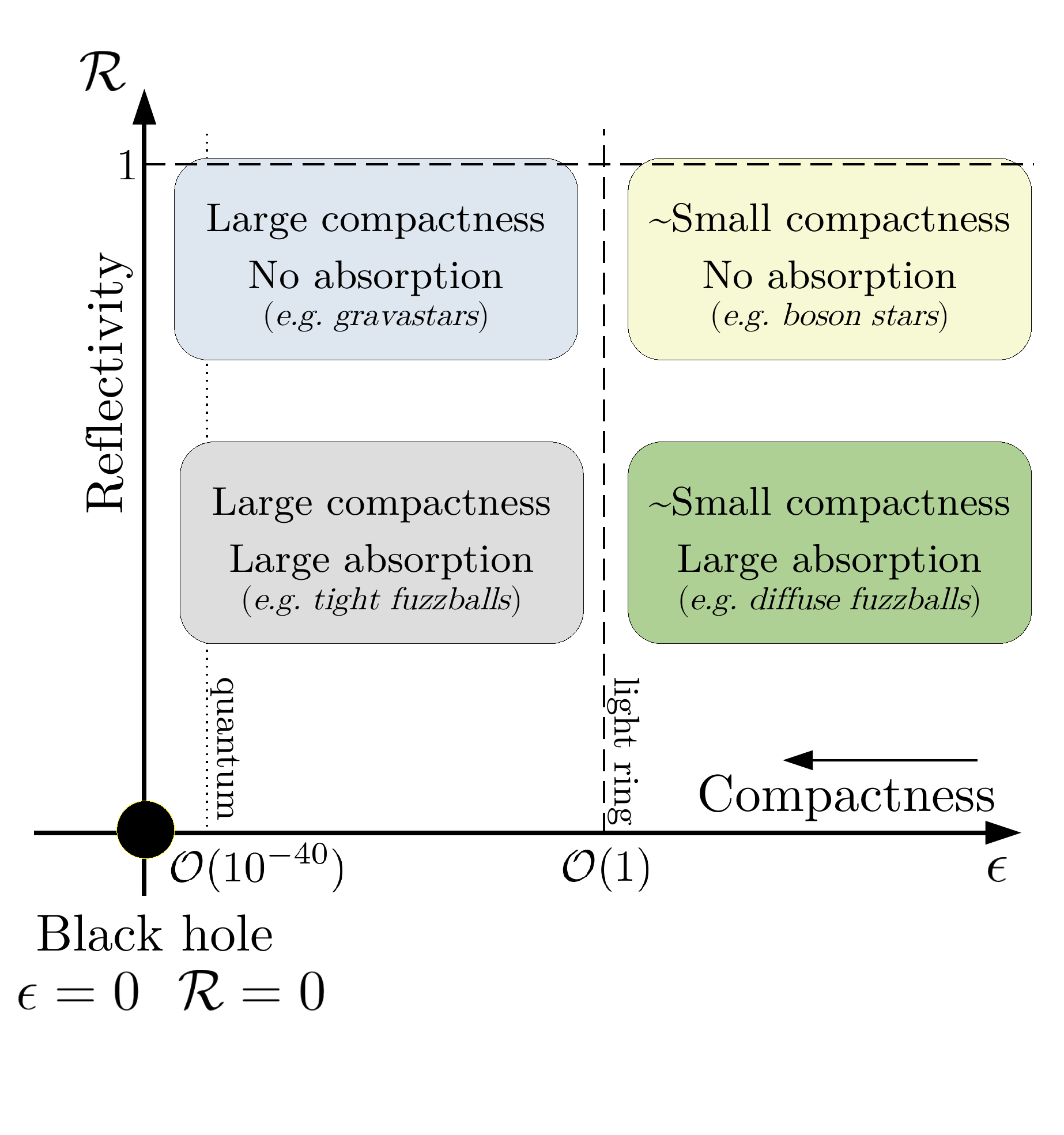}
\caption{Schematic representation of the parameter space of ECOs. The closeness parameter $\epsilon$ is related to the (effective) radius of the object by $r_0=r_+(1+\epsilon)$, where $r_+$ is the horizon location for a Kerr BH with same mass and spin. Models with quantum corrections at the horizon scale imply $r_0-r_+\approx \ell_{\rm Planck}$, hence $\epsilon={\cal O}(10^{-50}-10^{-40})$ for stellar to supermassive objects, depending on their mass. The reflectivity is related to the object's interior and is generically complex and frequency-dependent. For some reviews on different ECO models see~\cite{Carballo-Rubio:2018jzw,Cardoso:2019rvt}. For the distinction between tight and diffuse fuzzballs see~\cite{Guo:2017jmi}.} 
\label{fig:paramspace}
\end{figure}

Deviations from the BH hypothesis require either corrections to GR or beyond-standard model fields coupled to gravity (see \cite{Carballo-Rubio:2018jzw,Cardoso:2019rvt} for some reviews). Hypothetical dark compact objects without a classical BH horizon that, nonetheless, can mimic the phenomenology of BHs at the classical level are generically called ``BH mimickers"\cite{Cardoso:2007az,Lemos:2008cv} or exotic compact objects~(ECOs)~\cite{Giudice:2016zpa}. ECOs may be classified~\cite{Cardoso:2019rvt} in terms of their (see Fig.~\ref{fig:paramspace}) 
\begin{itemize}
    \item {\it compactness}, i.e. the inverse of their --~possibly effective~-- radius $r_0$ in units of the total mass $M$. It is customary to define a ``closeness" parameter $\epsilon$ such as $r_0=r_+(1+\epsilon)$, where $r_+$ is the location\footnote{For 
spherical objects the definition of the closeness parameter $\epsilon$ is coordinate-independent ($2\pi r_0$ is the proper equatorial circumference of the object). More in general one can directly relate $\epsilon$ to gauge-independent quantities. } of the horizon of the corresponding Kerr BH with the same mass and spin. Therefore, the BH limit corresponds to $\epsilon\to0$;
    \item {\it reflectivity} ${\cal R}$ at their (possibly effective) surface. This quantity is generically complex and frequency-dependent and can also depend on other object's parameters such as the spin. Note that ${\cal R}\to0$ in the limit of perfect absorption by a classical BH horizon.
\end{itemize}
Another mention-worthy property  used to classify ECOs is their so-called \emph{softness}, associated with the spacetime curvature at its surface. When the ECO's underlying theory is associated to a new length scale ${\cal L}\ll M$, the curvature (e.g. the Kretschmann scalar ${\cal K}$) at the surface can be much larger than the corresponding horizon curvature ${\cal K}\gg 1/M^4$. On the other hand, ECOs not motivated by new length scales other than $M$ (or parametrically close to it) cannot sustain larger curvatures at their surfaces. To the former we call ``hard'' ECOs while the latter are denoted by ``soft ECOs''.

A useful compass to navigate the ECO atlas is provided by the Buchdhal's theorem~\cite{Buchdahl:1959zz}, which states that, under certain assumptions, the maximum compactness of a self-gravitating object is 
$M/r_0=4/9$ (i.e., $\epsilon\geq 1/8$). This result prevents the 
existence of ECOs with compactness arbitrarily close to that of a BH. Relaxing some of these assumptions~\cite{Urbano:2018nrs} provides a way to circumvent the theorem and suggests a route to classify ECOs~\cite{Cardoso:2019rvt}.
In addition to some technical assumptions, Buchdahl's theorem assumes GR, spherical symmetry, and especially the fact that the matter sector is described by a single perfect fluid which is at most mildly anisotropic (tangential pressure smaller than the radial one).
Besides the assumption of GR (which is therefore violated in any modified-gravity theory), a quite common property of ECOs is the presence of an anisotropic pressure. Strong tangential stresses are necessary in many models to support very compact self-gravitating configurations. This is the case, for instance, of boson stars~\cite{Kaup:1968zz,Ruffini:1969qy}, gravastars~\cite{Mazur:2001fv,Mazur:2004fk,Mottola:2006ew,Cattoen:2005he}, ultracompact anisotropic stars~\cite{1974ApJ...188..657B,Raposo:2018rjn}, and wormholes~\cite{Morris:1988cz,Visser:1995cc,Lemos:2003jb}.

Although other classifications are possible, from a phenomenological perspective it is useful to divide ECOs into two classes: (i) ``ab initio" models that are solutions to consistent field theories coupled to gravity; (ii) phenomenological models that are studied to test possible generic implications of the absence of a horizon in dark compact objects, without a complete embedding in a concrete model. Examples of the first category are boson stars~\cite{Kaup:1968zz,Ruffini:1969qy,Liebling:2012fv} and fuzzballs~\cite{Lunin:2002qf,Mazur:2004fk, Mathur:2005zp, Mathur:2008nj,Mathur:2009hf,Mayerson:2020tpn}, whereas the second category comprises horizonless, possibly regular, phenomenological metrics that do not arise from a given theory, or that require ad-hoc matter fields in order for these metrics to solve Einstein's equations, as in the case of certain wormhole metrics.

The zoo of proposed ECO models is large and ever growing. It is not our scope to describe each model, for this we refer the interested reader to recent reviews~\cite{Carballo-Rubio:2018jzw,Cardoso:2019rvt}.
However, some ECO models stand out for being particularly interesting or representative of a generic class. In the following we shall discuss two examples of ``ab initio" models.

The most studied example of ECOs are {\it boson stars}, self-gravitating solutions formed by massive complex\footnote{In the case of real bosonic fields, similar self-gravitating solutions called \textit{oscillatons} exist in the same theories~\cite{Seidel:1993zk}. They have a weak time dependence and slowly decay, but can be very long lived and their phenomenology is similar to that of boson stars, whose metric is instead stationary.} bosonic fields, minimally coupled to GR~\cite{Kaup:1968zz,Ruffini:1969qy}. Their properties depends strongly on the bosonic self potential and various models with different classes of self-interactions~\cite{Colpi:1986ye,Schunck:2003kk,Kleihaus:2005me,Grandclement:2014msa,Minamitsuji:2018kof,Guerra:2019srj} and different field content~\cite{Brito:2015pxa,Herdeiro:2017fhv,Alcubierre:2018ahf} have been considered. Boson stars are the most robust model of ECOs, since they do not require modified gravity and their formation, stability, binary coalescence,  etc., can be studied from first principles~\cite{Liebling:2012fv,Palenzuela:2007dm,Palenzuela:2017kcg,Bezares:2017mzk,Sanchis-Gual:2018oui}. These ECOs are not meant to replace all BHs in the universe, for various reasons: indeed, just like ordinary neutron stars, their compactness is lower than the BH one, and equilibrium solutions have a maximum mass above which they are unstable against the gravitational collapse and classically form an ordinary BH. Finally, their mass scale is set by the mass of the bosonic field, which implies that a single bosonic field could give rise to boson stars only in a certain mass range. A family of boson stars ranging from (say) the stellar-mass to the supermassive range would require several different bosons with masses across several orders of magnitude.

More ambitious models of ECOs aim instead at replacing classical BHs completely, providing at the same time a quantum description of the horizon. A representative example is the {\it fuzzball proposal} of string theory~\cite{Lunin:2001jy, Lunin:2002qf, Mathur:2005zp, Mathur:2008nj}, wherein a classical BH is interpreted as an ensemble of regular, horizonless geometries that describes its quantum microstates~\cite{Myers:1997qi,Mathur:2005zp,Bena:2007kg,Balasubramanian:2008da,Bena:2013dka}. These geometries are solutions to (the low-energy truncations of) string theory --~hence circumvent Buchdhal's theorem~-- and have the same mass and charge of the corresponding BH. 
For special classes of extremal and charged BHs one can precisely count the microstates that account for the BH entropy~\cite{Strominger:1996sh, Horowitz:1996ay, Maldacena:1997de}, whereas in other cases the entropy counting is still an open problem.
In the fuzzball paradigm, all properties of the BH geometry emerge as an average over an ensemble of a large number of  microstates, or as a collective behavior of fuzzballs~\cite{Bianchi:2017sds, Bianchi:2018kzy, Bena:2018mpb, Bena:2019azk, Bianchi:2020des}, regardless of the curvature of the object~\cite{Mayerson:2020tpn}.

In the following we shall focus on the ECO phenomenology in a model-agnostic way, only occasionally referring to specific models.

\section{ECO Phenomenology}

GW astronomy allows us for unprecedented tests of the
nature of BHs and to search for new exotic species of compact objects. In this section we overview the main classes of GW-based tests, including inspiral tests of the multipolar structure and of the tidal deformability, ringdown tests, and searches for near-horizon structures with GW echoes.

\subsection{Multipole moments}

	Multipole moments were first introduced in the context of Newtonian mechanics (resp. electromagnetism) as a set of scalar quantities that appear on a multipolar expansion used to describe the gravitational (resp. electrostatic) potential $\Phi(\mathbf{x})$ resulting from a distribution of masses (resp. charges),  
\begin{equation}\label{eq:multipole_exp}
\Phi(\mathbf{x})=\sum_{\ell=0}^{\infty}\sum_{m=-\ell}^{\ell}\frac{M_{\ell m}}{r^{\ell+1}}\sqrt{\frac{4\pi}{2\ell +1}}Y_{\ell m}\left(\theta,\varphi\right)\,,
\end{equation}
where the expansion coefficients $M_{\ell m}$ are the \emph{mass multipole moments} of the body  (for the gravitational case, on which we shall focus in the following) and $Y_{\ell m}$ are the usual spherical harmonics. 
 In expansion~\eqref{eq:multipole_exp}, the multipole moments can in general be real or complex numbers. Nonetheless, if the potential $\Phi$ is real, the multipole moments must satisfy,
\begin{equation}
M_{\ell-m}=(-1)^m M_{\ell m}^{\,*}\,.
\end{equation}
 From the perspective of the interior of the object, these multipoles are related to the nonspherical distributions of matter within the body and they can be written as
%
\begin{equation}
M_{\ell m}=\sqrt{\frac{4\pi}{2\ell+1}}\int \rho(\mathbf{x})r^{\ell}Y^{\,*}_{\ell m} d^3 x \,.
\end{equation}
where $\rho$ is the mass density\footnote{The normalization of the multipole moments was chosen to coincide with the relativistic multipole moments (discussed below) when taking the Newtonian limit of the latter. Other work may use different normalization factors for the multipole moments, e.g.,~\cite{PoissonWill}.}. 

This simple Newtonian definition of multipole moments breaks down in GR due to the nonlinearity of Einstein's equations. Nonetheless, two independent formulations to define the relativistic multipole moments were developed, firstly by the works of Geroch and Hansen~\cite{Geroch:1970cd,Hansen:1974zz} and later by Thorne~\cite{Thorne:1980ru}. While the former presents an elegant mathematical definition of multipole moments, it is not suitable for computations for most astrophysical scenarios. Thus, henceforward we shall focus mostly on Thorne's approach. Remarkably, it was shown that the two definitions are equivalent~\cite{1983GReGr..15..737G}. 

In Thorne's formulation, the multipole moments of stationary and asymptotically flat spacetimes can be read-off directly from the metric as an extension to the procedure used to read the mass and the angular momentum from the asymptotic metric. Thorne's procedure requires the spacetime metric to be written in a specific class of coordinate systems called ``asymptotically Cartesian and mass centered''~(ACMC) where the metric approaches the asymptotically-flat Minkowski space sufficiently fast and the mass dipole term vanishes. This last condition is equivalent to set the origin of the coordinates at the center of mass of our system. In this ACMC form the metric reads, 
\begin{equation}
\label{eq:ACMC}
ds^2= dt^2 (-1+c_{00})+c_{0i}\, dt \, dx_i +(1+c_{00})\, dx_i^2	\,,
\end{equation}
with $c_{00}$ and $c_{0i}$ admitting a spherical-harmonic decomposition,
\begin{align}\label{cab_cart_coord}
c_{00} &{\,=\,} 2\sum_{\ell=0}^\infty \sum_{m=-\ell}^\ell \frac{1}{r^{1+\ell}}\sqrt{\frac{4\pi}{2\ell{\,+\,} 1}}  
\left( M_{\ell m} Y_{\ell m} {\,+\,} \ell'{\,<\,}\ell \right)\,,
\\
c_{0i} &{\,=\,} 2 \sum_{\ell=1}^\infty\sum_{m=-\ell}^{\ell}\frac{1}{r^{1+\ell}}\sqrt{\frac{4\pi (\ell{\,+\,} 1)}{ 
\ell(2\ell{\,+\,} 1)}}   
\left( S_{\ell m}Y^{B}_{i,\ell m}{\,+\,} \ell'{\,<\,}\ell \right) \,,\nonumber
 \end{align}
 where $Y^B_{i;\ell m}=\epsilon_{ij k} \,n_j\,r\partial_{k}Y_{\ell m}/{\sqrt{\ell(\ell+1)}}$ is the magnetic vector spherical harmonic, and $M_{\ell m}$ and $S_{\ell m}$ are the \emph{relativistic} mass and current multipole moments, respectively. In the Newtonian limit, the former reduces to well-known Newtonian mass multipole whereas the latter has no Newtonian analog. 

Remarkably, as long as the metric is written in ACMC coordinates the multipole moments are coordinate independent and  all coordinate dependence is pushed to the subleading terms in the metric.  In Eqs.~\eqref{cab_cart_coord} we already adopted the correct normalization factors to coincide with Geroch-Hansen multipole moments which are most commonly adopted throughout the literature~\cite{Cardoso:2016ryw}.

When the source of the gravitational field can be covered by so-called ``de Donder'' coordinates one can properly define the mass and current multipole moments using some ``effective'' mass and momentum densities, respectively~\cite{Thorne:1980ru}. Multipole moments of compact stars and ECOs can be computed using this method, however multipole moments of BHs can only be defined in terms of the external spacetime geometry.

\subsubsection{Testing the nature of compact objects with multipole moments}

A set of uniqueness and no-hair theorems predicts that, within GR, the outcome of the full gravitational collapse must be a Kerr BH~\cite{Carter:1971zc,Robinson:1975bv,Hawking:1973uf}. In spite of having an infinite multipolar structure, all the multipole moments of a BH can be related uniquely to only two parameters, its mass $M$ and the angular momentum $J$~\cite{Geroch:1970cd}. In the absence of rotation, BHs are spherically symmetric and the geometry reduces to the well-known  Schwarzschild metric, where the only nonvanishing multipole moment is the mass. The multipolar structure of a Kerr BH can be elegantly written as~\cite{Hansen:1974zz},
\begin{equation}
\label{eq:nohair}
M_{\ell}^{\rm BH}+ i S_\ell^{\rm BH}=M^{\ell+1}\left(i \chi\right)^\ell\,,
\end{equation}
where ${\chi:=J/M^2}$ is the dimensionless spin and 
\begin{equation}
 M=M_{0 0} \,,  \qquad  J=S_{1 0}\,,  \qquad   M_\ell=M_{\ell 0}\,, \qquad  S_\ell=S_{\ell 0}\,. 
\end{equation}

Since the Kerr metric is axisymmetric, only multipole moments with $m=0$ appear in Eq.~\eqref{eq:nohair}.  In addition, Kerr BHs have vanishing mass multipole moments when $\ell$ is odd and vanishing current multipole moments when $\ell$ is even. This is a consequence of the Kerr solution being \emph{axially} and \emph{equatorially symmetric}.

This elegant simplicity does not hold for other types of compact objects. There are no physical reasons for compact objects to have this same multipolar structure and nothing prevents them from being deformed even when non-rotating (e.g., there is no analog to Birkhoff's theorem beyond spherical symmetry). There is also no argument for the multipolar structure to be such as to satisfy the axial and the equatorial symmetry. In the most general scenario, it can be argued that no symmetries are expected for ECOs and their multipolar structure  will be rich and nontrivial.

In general, one can summarize this statement by parametrizing the multipole moments of an ECO as:
\begin{equation}
\label{eq:momentsECOs}
M^{\rm ECO}_{\ell m}=M^{\rm BH}_{\ell} + \delta M_{\ell m}\quad\,,\quad
S^{\rm ECO}_{\ell m}=S^{\rm BH}_{\ell} + \delta S_{\ell m}\,,
\end{equation}
where $\delta M_{\ell m}$ and $\delta S_{\ell m}$ are some model-dependent corrections to the mass and current multipole moments, whose value can be found by matching with the interior solution of the object or by other microphysical arguments. 

Generically, the most dominant smoking-gun signals of this ``non-Kerrness'' property are the current quadrupole moment $S_2$ (which breaks the equatorial symmetry of the Kerr solution) and a complete mass quadrupole tensor with three independent components $M_{2m}$ for $m=0,1,2$ (which breaks the axisymmetry of the Kerr solution). 

\begin{figure}[t]
\centering
\includegraphics[width=0.85\textwidth]{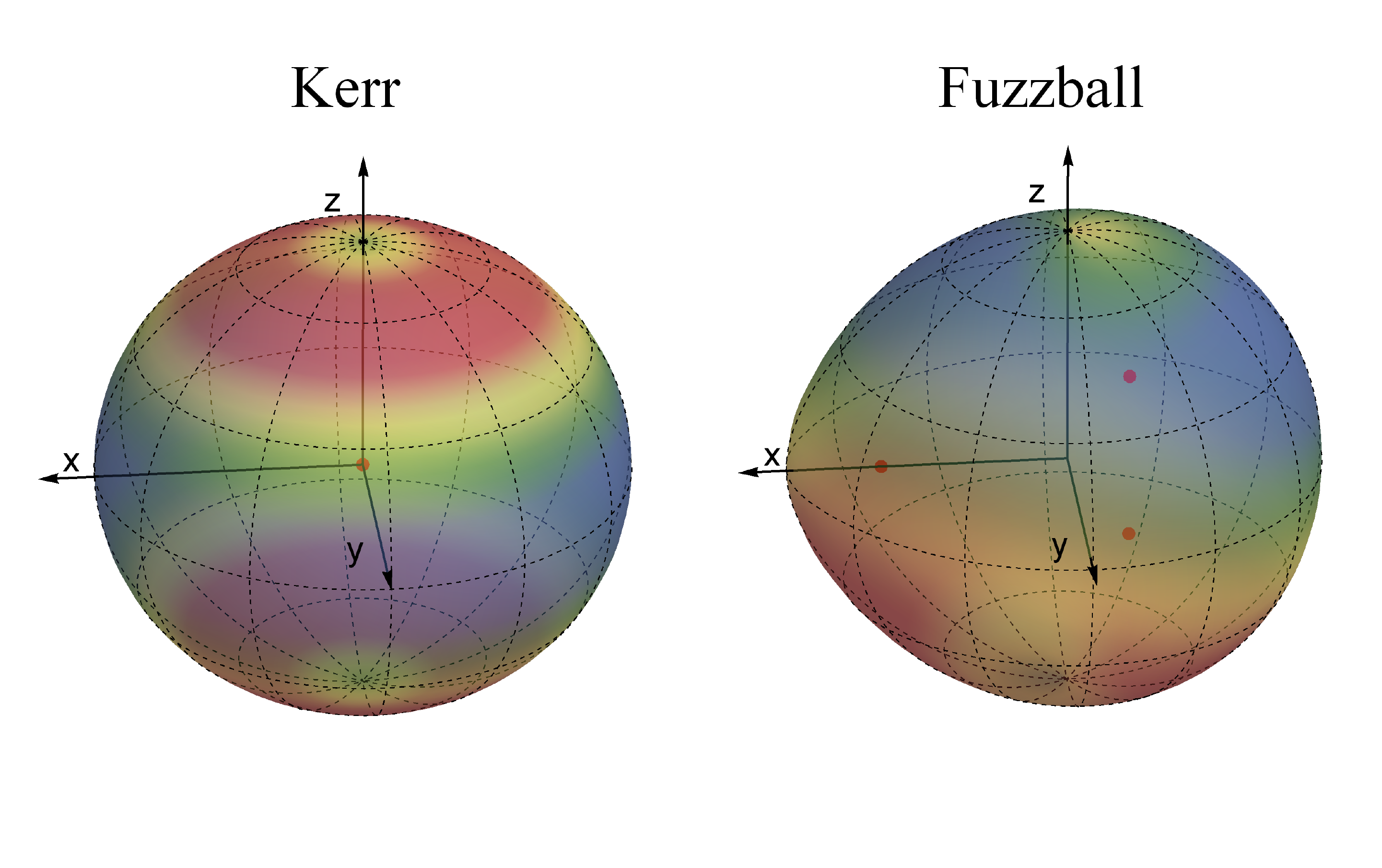}
\caption{Representation of the multipolar structure of different compact objects in isolation. The shape describes the embedding plot of constant $t$ and $r$ surfaces of the metric, while the colors were weighted according to $g_{t\phi}$ to represent the leading current multipole moments.  BHs (on the left) are described by the Kerr metric and have a unique multipolar structure. Nonrotating BHs are spherical, while rotating BHs  have an oblate shape. In addition, BHs have the remarkable feature that their moments are equatorially and axially symmetric. On the other hand, ECOs can have a richer multipolar structure. Fuzzballs (on the right), a string-theory motivated horizonless compact object which can be interpreted as a multicenter (represented by the red dots) extension of the BH model, can have a nontrivial multipolar structure and in general scenarios their multipole moments can break both the equatorial and axial symmetry~\cite{Bianchi:2020bxa,Bianchi:2020miz,raposothesis}. }
\label{fig:multipoles}
\end{figure}

Some particular ECO solutions provide specific examples of this complex multipolar structure (see Fig.~\ref{fig:multipoles}), as in the case of multipolar boson stars~\cite{Herdeiro:2020kvf} and fuzzball microstate geometries~\cite{Bena:2020see,Bianchi:2020bxa,Bena:2020uup,Bianchi:2020miz}.

 ``Soft'' ECOs motivated by some new physics effect whose new scale ${\cal L}$ is comparable to the mass (and for which the curvature at the surface is comparable to the corresponding horizon curvature) cannot have arbitrarily large deviations from the BH multipole moments. The ``softness'' property of these ECOs requires that the multipole moment deviations vanish in the BH limit sufficiently fast~\cite{Raposo:2018xkf}. The characteristic vanishing behaviour depends on the nature of the multipole moments. For axisymmetric spacetimes, spin-induced moments must vanish logarithmically (or faster), while non-spin induced moments vanish linearly (or faster), 
\begin{equation}
\frac{\delta M_{\ell}}{M^{\ell+1}}\to a_\ell\frac{\chi^\ell}{\log \epsilon}+b_\ell \epsilon\ldots
\end{equation}
and equivalently for the current multipole moments. In the expression above, $a_\ell$ and $b_\ell$ are numbers of order unity and the ellipsis stands for subleading terms. This particular behaviour suggests that any multipole moment detection will be dominated by the spin-induced quadrupole instead of any generic intrinsic quadrupole moment, unless the angular momentum is sufficiently low,
\begin{equation}
\chi\ll\sqrt{\epsilon|\log\epsilon|}    \,.
\end{equation}

The multipole moments can be measured through GW observations of binary coalescences. The multipole moments of compact objects in a binary system affect the point-particle phase of the emitted GW signal. The dominant term for this effect is the quadrupole moment $M_2$ which enters at 2PN order as~\cite{Krishnendu:2017shb},
\begin{equation}
\psi_{\ell=2}=\frac{75}{64}\frac{\left(m_2M_2^{(1)}+m_1M_2^{(2)}\right)}{\left(m_1m_2\right)^2}\frac{1}{v}\,,
\end{equation}
with $m_i$ the mass of the $i$-th body, $M_2^{(i)}$ its ($m=0$) quadrupole moment, and $v$ the orbital velocity. Putting constrains on the quadrupole provides a way to test the nature of compact objects already at 2PN-level.
Constraints on parametrized PN deviations using GW events~\cite{LIGOScientific:2018mvr,LIGOScientific:2019fpa,Abbott:2020jks} can be mapped into a constraint on $\delta M_{20}$, in particular using light binaries which perform many cycles in band before merger. However, such tests are challenging due to the fact that the $2$PN term in the GW phase depends also on the binary component spins, which have not been measured accurately so far. This introduces correlations between the spin and the quadrupole moment.

While current GW constraints will become (slightly) more stringent in the next years as the sensitivity of the 
ground-based detectors improve~\cite{Krishnendu:2018nqa,Krishnendu:2019ebd,Kastha:2018bcr,Kastha:2019brk}, much tighter bounds will come from extreme mass-ratio 
inspirals~(EMRIs), one of the main targets of the future space mission LISA~\cite{Audley:2017drz}.  Although EMRI data 
analysis is challenging~\cite{Babak:2017tow,Chua:2018yng,Chua:2019wgs,LISADataChallenge},
a detection of 
these systems can be potentially used to measure the ($m=0$, mass) quadrupole moment $\delta{M}_2$ of the central 
supermassive object with an accuracy of one part in $10^4$~\cite{Barack:2006pq,Babak:2017tow} and of a large set of high-order multipole moments~\cite{Kastha:2019brk}, offering unprecedented 
tests of the nature of supermassive objects~\cite{Glampedakis:2005cf,Raposo:2018xkf,Destounis:2020kss}.

\subsection{Tidal heating}

If the binary's components are dissipative systems, energy and angular momentum will be dissipated in their interior in addition to the GW emission to infinity. For BHs, energy and angular momentum absorption by the horizon is responsible for tidal heating~\cite{Hartle:1973zz,PhysRevD.64.064004}. This effect is particularly significant for highly spinning BHs and in the latest stages of the inspiral, since it enters the GW phase at $2.5{\rm PN}\times \log v$ order ($4{\rm PN}\times \log v$ order) for spinning (nonspinning) binaries. In the absence of the $\log v$ term, such corrections would be completely degenerate with the time and phase of coalescence in the waveform, and therefore unmeasurable. They anyway remain mildly correlated with other parameters and therefore hard to measure.
For LISA binaries, constraints of the amount of dissipation would be stronger for highly spinning objects and for binaries with large mass ratios~\cite{Maselli:2017cmm,Datta:2019epe}.

On the other hand, tidal heating can contribute to thousands of radians of accumulated orbital phase for EMRIs in the LISA band~\cite{Hughes:2001jr,Bernuzzi:2012ku,Taracchini:2013wfa,Harms:2014dqa,Datta:2019euh,Datta:2019epe}. This would allow to distinguish between binary BHs and binary involving other compact objects (for which tidal dissipation is often negligible).  For EMRIs in the LISA band, this effect could be used to put a very stringent upper bound on the reflectivity of ECOs, at the level of $0.01\%$~\cite{Datta:2019epe}.

\subsection{Tidal deformability}

As the two compact objects in an inspiraling binary approach each others, tidal effects become increasingly relevant. The gravitational field of each object acts as a tidal field on its companion, deforming its shape and inducing some multipolar deformation in the spacetime. This effect can be quantified in terms of ``tidal-induced multipole moments''. A weak tidal field can be decomposed into the electric (or polar) tidal field moments ${\cal E}_{\ell m}$ and the magnetic (or axial) tidal field moments ${\cal B}_{\ell m}$. In the nonrotating case, the ratio between the multipole moments and the tidal field moments that induces them defines the tidal deformability of the body,
\begin{equation}
\lambda_E^{(\ell)}=\frac{M_{\ell m}}{{\cal E}_{\ell m}}\quad,\quad \lambda_B^{(\ell)}=\frac{S_{\ell m}}{{\cal B}_{\ell m}}\,.    
\end{equation}
These quantities are independent of $m$ and of the external tidal field (if the latter is sufficiently weak). 
It is convenient to introduce the dimensionless tidal Love numbers (TLNs) $k_\ell^{E}$ and $k_\ell^B$,
\begin{equation}
k_\ell^E={\rm const}\frac{\lambda_E^{(\ell)}}{M^{2\ell +1}}\quad,\quad k_\ell^B={\rm const}\frac{\lambda_B^{(\ell)}}{M^{2\ell +1}}\,,
\end{equation}
where different (dimensionaless) constant prefactors have been used in the literature, depending on the adopted conventions.

When the object is rotating, the angular momentum couples with the electric (resp. magnetic) tidal field moments to induce current (resp. mass) multipole moments on the body according a set of selection rules~\cite{Poisson:2014gka,Pani:2015hfa,Landry:2015zfa,Pani:2015nua}. This effect allows us to define some new classes of ``rotational TLNs'' for rotating objects, which have no Newtonian analog. 

A remarkable result in GR is that the TLNs of BHs are precisely zero. This was first demonstrated for nonrotating BHs~\cite{Binnington:2009bb,Damour:2009vw,Gurlebeck:2015xpa} and then extended for slowly rotating BHs~\cite{Poisson:2014gka,Pani:2015hfa,Landry:2015zfa}, but more recently it has been extended to Kerr BHs without any approximations~\cite{Chia:2020yla} (see also Refs.~\cite{LeTiec:2020spy,LeTiec:2020bos}).  This characteristic behavior is associated to the regularity conditions for tidal perturbations at the horizon. Thus, due to the absence of a horizon, the TLNs of ECOs will in general be nonzero and in principle can provide a smoking-gun test of the nature of dark ultracompact objects~\cite{Cardoso:2017cfl}.  

\begin{figure}[t]
\centering
\includegraphics[width=0.49\textwidth]{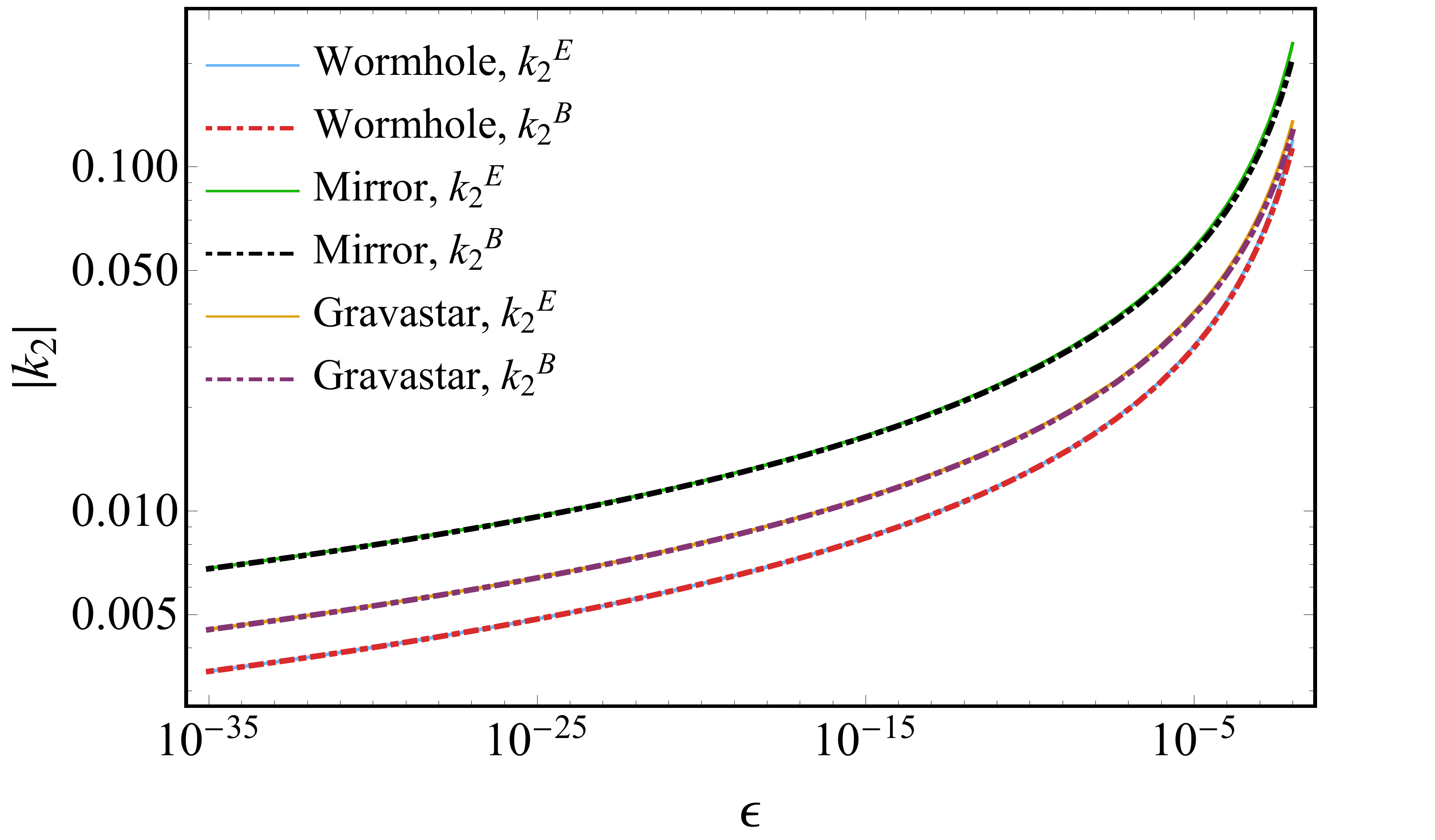}
\includegraphics[width=0.49\textwidth]{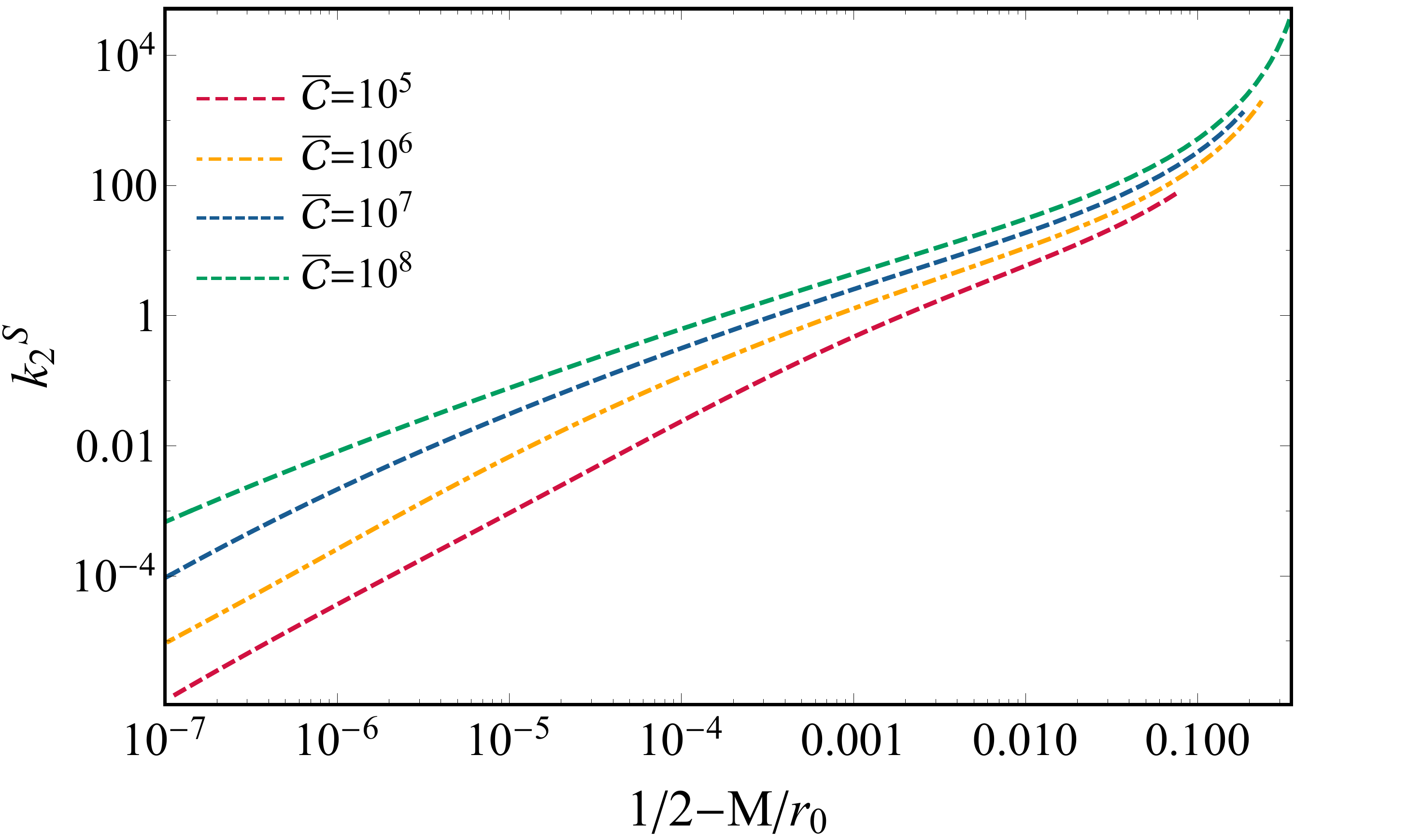}
\caption{TLNs of ``stiff'' ECOs (left panel)~\cite{Cardoso:2017cfl} and (scalar) TLNs of an anisotropic star model dependent on the anisotropic scale (right panel)~\cite{Raposo:2018rjn}. The latter depends on a parameter  $\bar{\cal C}$ that controls the anisotropy scale of the model (the isotropic limit is obtained by taking $\bar{\cal C}\to 0$).  The behavior of the TLNs depends on the particular boundary conditions at the surface. ECOs with  Robin-type boundary conditions at the surface (left) show a logarithmic vanishing behavior in the BH limit, while the TLNs of anisotropic stars vanish  with an approximately polynomial relation in this regime (right). Note that in the former case the axial and polar TLNs coincide as $\epsilon\to0$.} 
\label{fig:Love_ani}
\end{figure}

The TLNs were explicitly computed for different models of ECOs such as boson stars~\cite{Cardoso:2017cfl,Sennett:2017etc,Mendes:2016vdr}, gravastars~\cite{Pani:2015tga,Cardoso:2017cfl,Uchikata:2016qku}, anisotropic stars~\cite{Raposo:2018rjn} and other simple ECOs with stiff EOS at the surface~\cite{Cardoso:2017cfl}. As expected, it was found that the TLNs are generically nonzero and vanishing in the BH limit. For the particular class of ``stiff" ECOs, i.e with  Robin-type boundary conditions at the surface~\cite{Maselli:2018fay} (e.g. due to a perfectly reflective surface) it was found that the TLNs exhibit a logarithmically vanishing behavior in the BH limit~\cite{Cardoso:2017cfl},
\begin{equation}
k_\ell^{\rm ECO}\to\frac{a_\ell}{1+b_\ell\log(\epsilon)}\,,\qquad \epsilon\to0
\end{equation}
where $a_\ell$ and $b_\ell$ are some model-dependent $\sim{\cal O}(0.01-1)$ dimensionless constants. This remarkable logarithmic dependence suggests that, although small, the TLNs are not infinitesimally vanishing even for ultracompact objects. For spherical objects with surfaces located at Planckian distances from the corresponding horizon location, this magnifying effect yields $k_2\sim {\cal O}(10^{-3})$. In Fig.~\ref{fig:Love_ani} we show the leading (electric and magnetic) TLNs for these models of ECOs. In addition to this behavior, we see that the TLNs of this class of ECOs exhibit an interesting isospectrality in the BH limit, i.e, polar and axial Love numbers coincide for ultracompact configurations ($\epsilon\to0$). 
In contrast, other ECOs such as anisotropic stars with a ``smoother'' transition between the vacuum external region and the interior fluid matter, show some approximately polynomial vanishing behavior in the BH limit~\cite{Raposo:2018rjn},
\begin{equation}
    k_\ell^{\rm ECO}\to a_\ell \left(\frac{\epsilon}{M}\right)^n\,,
\end{equation}
where $a_\ell$ and $n$ are some model dependent parameters. 

Similarly to the effect of multipole moments discussed above, the effect of tidal deformability alters the GW signal of a compact object inspiral by adding a $5$PN correction to the GW phase,
\begin{equation}
\psi_{\rm TD}=-\psi_{\rm N}\frac{624\Lambda}{m^5}v^{10},
\end{equation}
where $\Lambda$ is the weighted tidal deformability that contains the contribution of the $\ell=2$ polar tidal deformability of both compact objects in the binary. Other TLNs enter the waveform at even higher order.

Current (and especially future) GW interferometers could measure the TLNs of BH mimicker in order to distinguish it from a BH~\cite{Cardoso:2017cfl,Sennett:2017etc,Maselli:2017cmm}. In the comparable-mass case, a precise measurement requires highly-spinning supermassive ECO binaries up to $10\,{\rm Gpc}$ detectable by LISA.
LISA may also be able to perform model selection between different families of BH mimickers \cite{Maselli:2018fay}, 
 although this will in general require detection of 
 golden binaries, with very large signal-to-noise ratio~\cite{Addazi:2018uhd}.
Finally, EMRI observations can set even more stringent constraints, since the measurement errors on the Love number scale as $q^{1/2}$, where $q\ll 1$ is the mass ratio of the binary, potentially constraining the Love number of the central object within one part in $10^5$~\cite{Pani:2019cyc}.

Typically the effects discussed above are included independently in inspiral waveform templates. However, for concrete models like boson stars it is possible to consistently include several corrections (multipolar structure, tidal heating, TLNs) in the inspiral signal, improving  the accuracy on the measurement of the (fewer) free parameters of the template~\cite{Pacilio:2020jza}.

\subsection{Ringdown}
The ringdown is the final stage of a compact binary coalescence when the remnant relaxes to an equilibrium solution. When the remnant is a BH, the ringdown is dominated by its complex characteristic frequencies, the so-called quasi-normal modes~(QNMs), which describe the response of the compact object to a perturbation~\cite{Teukolsky:1973ha,Press:1973zz,1974ApJ...193..443T,Chandrasekhar:1975zza,Leaver:1985ax,Kokkotas:1999bd,Berti:2009kk}. In the linear regime, the spacetime metric can be written as $g_{\mu \nu} = g_{\mu \nu}^{(0)} + h_{\mu \nu}$, where $g_{\mu \nu}^{(0)}$ is the background metric of the compact object and $h_{\mu \nu} \ll 1$ is the perturbation.
The BH ringdown signal can be modeled as a linear superposition of exponentially damped sinusoids
\begin{equation}
    h = \sum_{\ell mn} A_{\ell mn}(r) e^{-t/\tau_{\ell mn}} \sin(\omega_{\ell mn} t + \phi_{\ell mn}) ~_{-2}\mathcal{Y}_{\ell m}(\theta,\varphi) \,,
\end{equation}
where $\omega_{\ell mn}$ are the characteristic frequencies of the remnant, $\tau_{\ell mn}$ are the damping times, $A_{\ell mn}(r)\propto 1/r$ is the amplitude of the signal at a distance $r$, $\phi_{\ell mn}$ is the phase, and $~_{s}{\cal Y}_{\ell m}(\theta,\varphi)\propto e^{i m\varphi}$ are the spin-weighted spheroidal harmonics which depend on the location of the observer with respect to the source. Each mode is described by three integers, namely the angular number ($\ell \geq 0$), the azimuthal number $m$ (such that $|m| \leq \ell$), and the overtone number ($n \geq 0$).

From the detection of the ringdown it is possible to infer the QNMs of the remnant and understand the nature of compact objects. Due to the GR uniqueness theorem, the QNM spectrum of Kerr BHs depends uniquely on two parameters, i.e., their mass and angular momentum~\cite{Carter71,Robinson:1975bv}. As a consequence, a test of the Kerr hypothesis would require the identification of at least two QNMs, whereas a detection of several modes would allow for multiple independent tests of the null-hypothesis.

Up to date, the least-damped QNM ($\ell=m=2$, $n=0$) has been observed in the ringdown of several GW events and is compatible with a Kerr BH remnant with the mass and the spin predicted by the inspiral~\cite{TheLIGOScientific:2016src,LIGOScientific:2019fpa,Abbott:2020jks,Ghosh:2021mrv}. Few loud GW signals show evidence for the first overtone ($\ell=m=2$,  $n=1$) whose frequency measurements could set constraints on GR~\cite{Giesler:2019uxc,Isi:2019aib} (but see \cite{Bhagwat:2019dtm,Ota:2019bzl,Forteza:2020hbw} for a related discussion), whereas the damping times are less constrained~\cite{Abbott:2020jks,Ghosh:2021mrv}. Third generation ground-based detectors and the future LISA mission will allow for unprecedented tests of the BH paradigm given the expected large signal-to-noise ratio in the ringdown~\cite{Audley:2017drz,Berti:2016lat,Maggiore:2019uih,Reitze:2019iox}.

\begin{figure}[t]
\centering
\includegraphics[width=0.7\textwidth]{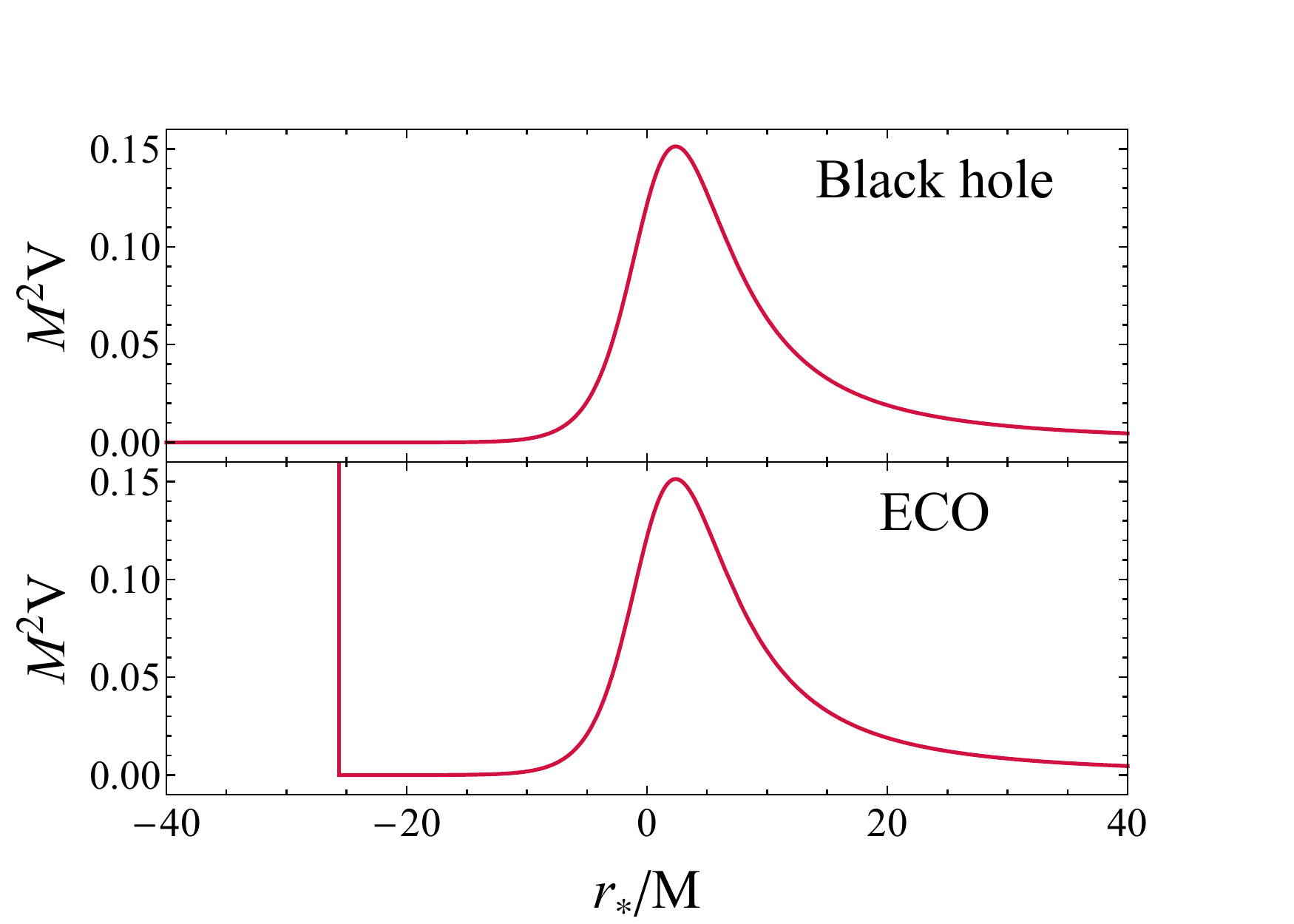}
\caption{Effective potential for axial gravitational perturbations of a Schwarzschild BH (top panel) and an ECO with $\epsilon=10^{-6}$ (bottom panel). The effective potentials have a maximum approximately at the photon sphere, $r \approx 3M$. In the ECO case, the effective potential features a cavity between the radius of the ECO and the photon sphere~\cite{Cardoso:2016rao,Cardoso:2016oxy,Cardoso:2019rvt}.} 
\label{fig:effectivepotential}
\end{figure}

ECOs can be distinguished from BHs from their different linear response. For simplicity, let us analyse a static ECO whose exterior spacetime, assuming GR as a reliable approximation, is described by the Schwarzschild metric (see Ref.~\cite{Maggio:2018ivz} for the extension to the spinning case)
\begin{equation}
 ds^2 = -f(r) dt^2 + \frac{1}{f(r)} dr^2 + r^2 (d\theta^2 + \sin^2 \theta d\varphi^2) \,,    
\end{equation}
where $f(r)=1-2M/r$ and $M$ is the total mass of the compact object.
In order to derive the QNM spectrum of the ECO, let us perturb it with a spin-$s$ perturbation where $s=0,\pm1,\pm2$ for scalar, electromagnetic and gravitational perturbations, respectively. The perturbation can be decomposed as
\begin{equation}
    \Psi_s(t,r,\theta,\varphi) =  \sum_{\ell m}  ~_{s}{\cal Y}_{\ell m}(\theta,\varphi) ~_{s}\psi_{\ell m}(r) e^{-i \omega t}\,, \label{psi}
\end{equation}
where in the following we will omit the $s,\ell,m$ subscripts for brevity. The radial component of the perturbation is governed by a Schr\"odinger-like equation~\cite{Regge:1957td,Zerilli:1970se,Zerilli:1971wd}
\begin{equation}
    \frac{d^2 \psi(r)}{dr_*^2} + \left[ \omega^2 - V(r)\right] \psi(r) = 0 \,, \label{waveeq}
\end{equation}
where $r_*$ is the tortoise coordinate
\begin{equation}
  r_* = r+2M \log \left( \frac{r}{2M} -1\right)  \,,
\end{equation}
and the effective potential is
\begin{eqnarray}
    V_{\rm axial} &=& f \left( \frac{\ell(\ell+1)}{r^2} + (1-s^2)\frac{2M}{r^3}\right) \,, \label{Vaxial} \\
    V_{\rm polar} &=& 2f \left( \frac{q^2 (q+1) r^3 + 3 q^2  M r^2 + 9 M^2 (q r + M)}{r^3 (qr+3M)^2} \right) \,, \label{Vpolar}
\end{eqnarray}
where $q=(\ell-1)(\ell+2)/2$. The potential in Eq.~\eqref{Vaxial} describes scalar, electromagnetic and axial gravitational perturbations, whereas the potential in Eq.~\eqref{Vpolar} describes polar gravitational perturbations. The effective potential as a function of the tortoise coordinate is shown in Fig.~\ref{fig:effectivepotential} for axial gravitational perturbations. It has a maximum approximately at the photon sphere $r\approx3M$, which is the unstable circular orbit of photons around the compact object. In the ECO case, the absence of the event horizon at $r_* \to -\infty$ implies the existence of a cavity between the ECO radius and the photon sphere. The cavity can support long-lived trapped modes which are responsible for a completely different QNM spectrum with respect to the BH case. 

\begin{figure}[t]
\centering
\includegraphics[width=0.7\textwidth]{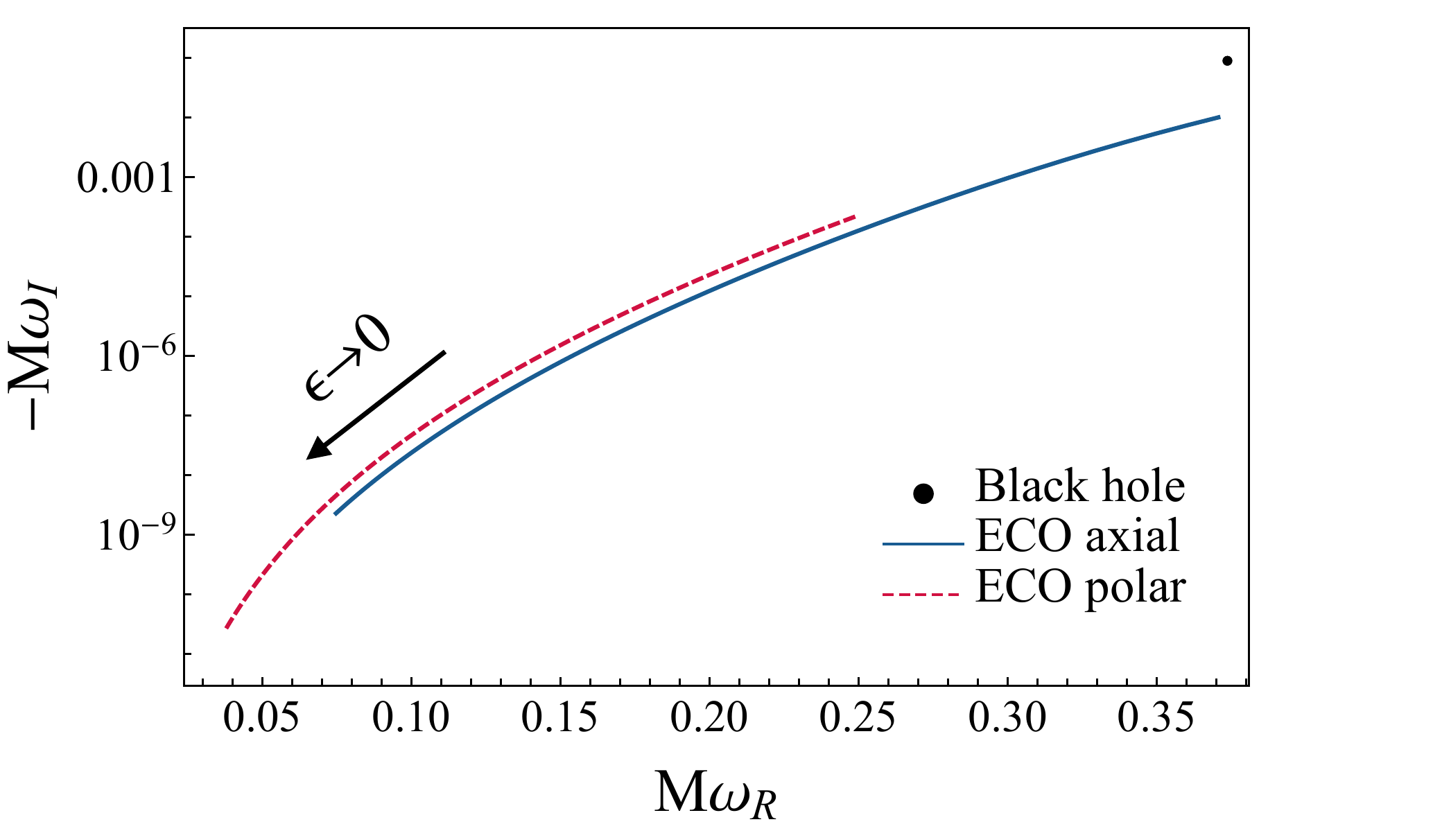}
\caption{QNM spectrum of a perfectly reflecting ECO with $\epsilon \in (10^{-2},10^{-10})$ compared to the fundamental $\ell=2$ gravitational QNM of a Schwarzschild BH. Axial and polar modes are not isospectral at variance with the BH case. As $\epsilon \to 0$, the ECO QNMs are low-frequencies and long-lived~\cite{Cardoso:2016rao}. } 
\label{fig:lowfrequencyQNMs}
\end{figure}

In order to derive the QNMs of ECOs, let us impose boundary conditions at infinity and at the radius of the ECO. Indeed, Eq.~\eqref{waveeq} with the addition of two boundary conditions defines an eigenvalue problem whose complex eigenvalues are the QNMs of the object, $\omega=\omega_R + i \omega_I$. According to the convention in Eq.~\eqref{psi}, a stable mode has $\omega_I < 0$, whereas an unstable mode has $\omega_I > 0$, with damping (instability) timescale $\tau_{\rm damping \ (inst)} \equiv 1/|\omega_I|$ for the former (latter) case. At infinity we impose that the perturbation is a purely outgoing wave 
\begin{equation}
    \psi \sim e^{i \omega r_*} \,, \quad \text{as} \ r_* \to + \infty \,.
\end{equation}
In the BH case, the event horizon would require that the perturbation is a purely ingoing wave as $r_* \to -\infty$ ($r \to r_+ = 2M$). In the ECO case, the solution is generally a superposition of ingoing and outgoing waves,
\begin{equation}
    \psi \sim C_{\rm in} e^{-i \omega r_*} + C_{\rm out} e^{i \omega r_*} \,, \quad \text{as} \ r_* \to r_*^0 \,, \label{solr0}
\end{equation}
where $r_*^0 \equiv r_*(r_0)$ and $r_0=2M(1+\epsilon)$. The surface reflectivity of the ECO is then defined as~\cite{Maggio:2017ivp}
\begin{equation}
    \mathcal{R} = \frac{C_{\rm out}}{C_{\rm in}} e^{2 i \omega r_*^0} \,.
\end{equation}
A perfectly reflecting ECO has $|\mathcal{R}|^2=1$, whereas a totally absorbing compact object has $|\mathcal{R}|^2=0$ as in the BH case. The boundary conditions that describe a perfectly reflecting ECO are~\cite{Maggio:2017ivp,Maggio:2018ivz}
\begin{eqnarray}
    \psi(r_0) &=& 0 \quad \rm Dirichlet \ on \ axial \,, \label{dir} \\
    d\psi(r_0)/dr_* &=& 0 \quad \rm Neumann \ on \ polar \,, \label{neu}
\end{eqnarray}
where for the Dirichlet (Neumann) boundary condition the waves are totally reflected with inverted phase $\mathcal{R}=-1$ (in phase $\mathcal{R}=1$). Fig.~\ref{fig:lowfrequencyQNMs} shows the QNM spectrum of a perfectly reflecting ECO compared to the fundamental $\ell=2$ gravitational QNM of a Schwarzschild BH, where $\epsilon \in (10^{-2},10^{-10})$ from the right to left of the plot. As shown in Fig.~\ref{fig:lowfrequencyQNMs}, an important feature of ECOs is the breaking of isospectrality between axial and polar modes unlike BHs in GR~\cite{Maggio:2020jml,Cardoso:2019mqo,Berti:2009kk}. Furthermore as $\epsilon \to 0$, the deviations from the BH QNM are arbitrarily large and the QNMs are low frequencies ($M \omega_R \ll 1$) and long-lived ($\tau_{\rm damping} \gg 1$)~\cite{Cardoso:2016rao}.

For $\epsilon \ll 1$, the QNMs can be derived analytically in the low-frequency regime~\cite{Vilenkin:1978uc,Maggio:2018ivz,Cardoso:2019rvt}
\begin{eqnarray}
M \omega_R &\simeq& -\frac{M \pi}{2 |r_*^0| } \left( q + \frac{s(s+1)}{2}\right)\,, \label{MwR}\\
M \omega_I &\simeq& -\beta_{s\ell} \frac{M}{|r_*^0|} (2M \omega_R)^{2\ell+2} \,, \label{MwI}
\end{eqnarray}
where $q$ is a positive odd (even) integer for polar (axial) modes and $\beta_{s\ell} = \left[\frac{(\ell-s)!(\ell+s)!}{(2\ell)!(2\ell+1)!!}\right]^2$ \cite{Starobinskij2}. Low-frequency modes are trapped in the cavity of the effective potential between the ECO radius and the photon sphere barrier. The real part of the QNMs scales with the width of the cavity as $\omega_R \sim |\log \epsilon|^{-1}$, whereas the imaginary part of the QNMs depends on the tunneling probability through the potential and scales as $\omega_I \sim - |\log \epsilon|^{-(2l+3)}$.

We note that Eq.~\eqref{solr0} is valid only in a region where the effective potential vanishes, so it requires $\epsilon\ll1$.  In order to define the ECO boundary condition more generally, one can make use of the BH membrane paradigm and generalize it to the case of horizonless compact objects. According to the BH membrane paradigm, a static observer outside the horizon can replace the interior of a perturbed BH by a fictitious membrane located at the horizon~\cite{Damour:1982,MembraneParadigm,Price:1986yy}. The properties of the membrane are fixed by the Israel-Darmois junction conditions~\cite{Darmois:1927,Israel:1966rt}
\begin{equation} \label{junction}
    [[K_{ab} - K h_{ab}]]=-8 \pi T_{ab} \,, \quad [[h_{ab}]]=0 \,,
\end{equation}
where $h_{ab}$ is the induced metric on the membrane, $K_{ab}$ is the extrinsic curvature, $K=K_{ab} h^{ab}$, $T_{ab}$ is the stress-energy tensor of the membrane and $[[...]]$ is the jump of a quantity at the membrane. The fictitious membrane is such that the extrinsic curvature of the interior spacetime vanishes. Consequently the junction conditions impose that the fictitious membrane is a viscous fluid with the stress-energy tensor~\cite{MembraneParadigm}
\begin{equation} \label{stressenergytensor}
    T_{ab} = \rho u_a u_b + (p- \zeta \Theta) \gamma_{ab} -2 \eta \sigma_{ab} \,,
\end{equation}
where $\eta$ and $\zeta$ are the shear and the bulk viscosities of the  fluid, $\rho$, $p$ and $u_a$ are the density, the pressure and the 3-velocity of the fluid, $\Theta=u^a_{;a}$ is the expansion, $\sigma_{ab}=\frac{1}{2} \left(u_{a;c} \gamma_b^c + u_{b;c} \gamma_a^c - \Theta \gamma_{ab}\right)$ is the shear tensor, $\gamma_{ab}=h_{ab} + u_a u_b$ is the projector tensor, and the semicolon is the covariant derivative compatible with the induced metric, respectively.

The BH membrane paradigm allows to describe the interior of a perturbed BH in terms of the shear and bulk viscosities of a fictitious fluid located at the event horizon. The generalization of the BH membrane paradigm to horizonless compact objects allows to analyze several models of ECOs with different internal structures in terms of the properties of a fictitious membrane located at the ECO radius~\cite{Maggio:2020jml,Abedi:2020ujo}. The shear and the bulk viscosities are generically complex and frequency-dependent and are related to the reflective properties of the ECO. In particular, for each model of ECO, the shear and the bulk viscosities are uniquely determined. In the following we shall focus on models of ECOs which are described by a Schwarzschild exterior.

The junction conditions in Eq.~\eqref{junction} with the stress-energy tensor in Eq.~\eqref{stressenergytensor} allow to derive generally the boundary conditions at the ECO radius (for details on the derivation see Ref.~\cite{Maggio:2020jml})
\begin{eqnarray}
    \frac{d\psi(r_0)/dr_*}{\psi(r_0)} &=& - \frac{i \omega}{16 \pi \eta} - \frac{r_0^2 V_{\rm axial}(r_0)}{2 (r_0-3M)} \,, \qquad \ \ \rm axial \,, \label{BCax} \\
    \frac{d\psi(r_0)/dr_*}{\psi(r_0)} &=& - 16 \pi i \eta \omega + F(r_0, \omega, \eta, \zeta) \,, \quad \rm polar \,, \label{BCpol}
\end{eqnarray}
where $F(r_0, \omega, \eta, \zeta)$ is a cumbersome function given in Ref.~\cite{Maggio:2020jml}. Let us notice that in the BH limit the boundary conditions in Eqs.~\eqref{BCax},~\eqref{BCpol} reduce to the BH case. Indeed, according to the BH membrane paradigm,
\begin{equation}
    \eta_{\rm BH} = \frac{1}{16 \pi} \,.
\end{equation}
For $\eta \to \eta_{\rm BH}$ and $r_0 \to 2M$ ($\epsilon\to0$), both the axial and polar boundary conditions in Eqs.~\eqref{BCax} and~\eqref{BCpol} describe a purely ingoing wave as in the BH case. Moreover for $\eta=0$ and $\epsilon \ll 1$, the boundary conditions in Eqs.~\eqref{BCax},~\eqref{BCpol} reduce to Dirichlet and Neumann boundary conditions on axial and polar modes, respectively, as in Eqs.~\eqref{dir} and~\eqref{neu}, thus describe a perfectly reflecting ECO. The case of a partially absorbing surface is analyzed by considering $\eta \in (0,\eta_{\rm BH})$.
Interestingly as the ECO radius approaches the photon sphere, $r_0 \to 3M$, the axial boundary condition reduces to $\psi(r_0)=0$ for any $\eta \in \mathbb{C}$. As a consequence, an ECO with $r_0=3M$ is a perfect reflector of axial modes regardless of its interior structure. The same universality does not occur in the polar sector.

\begin{figure}[t]
\centering
\includegraphics[width=0.49\textwidth]{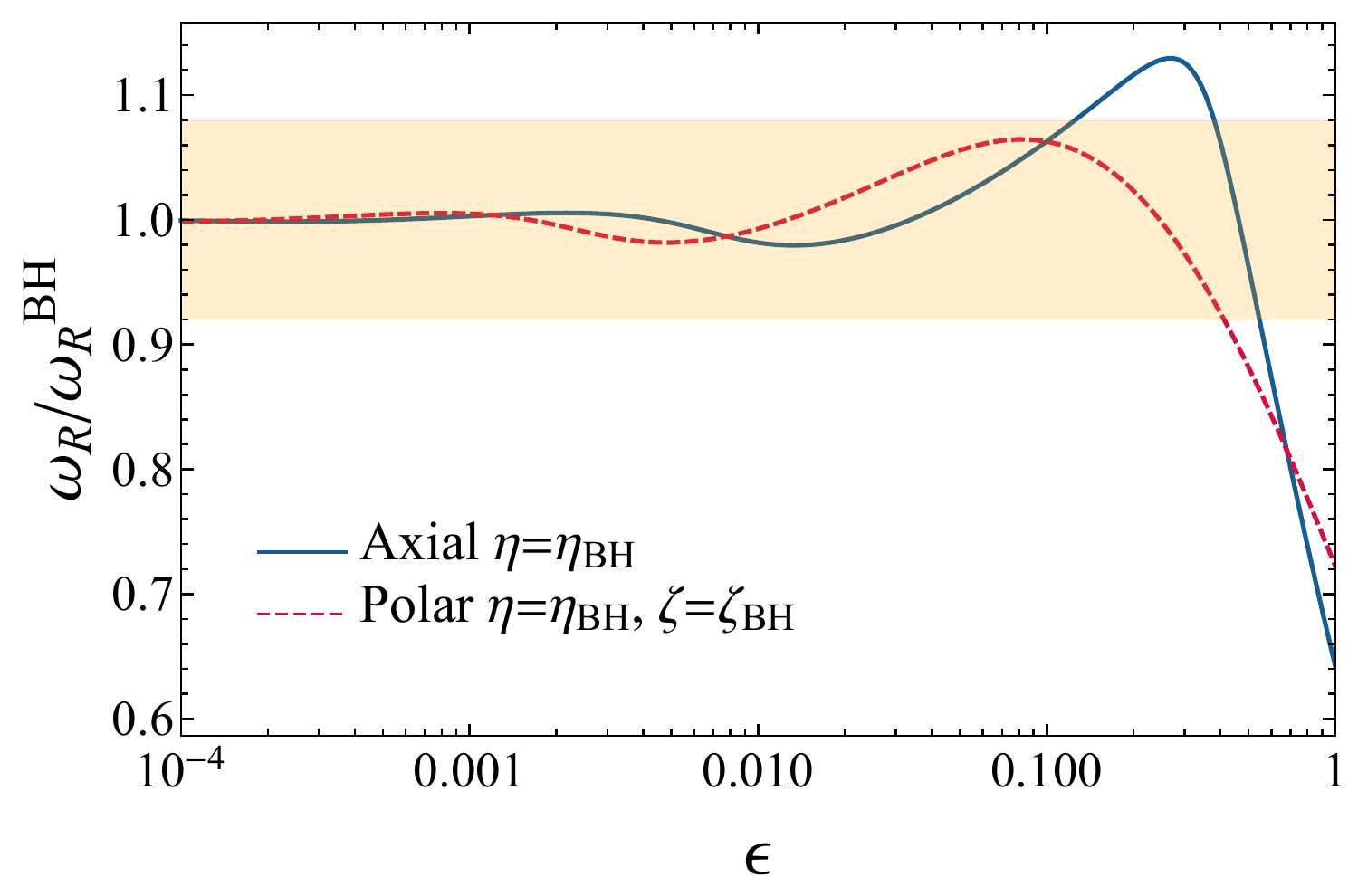}
\includegraphics[width=0.49\textwidth]{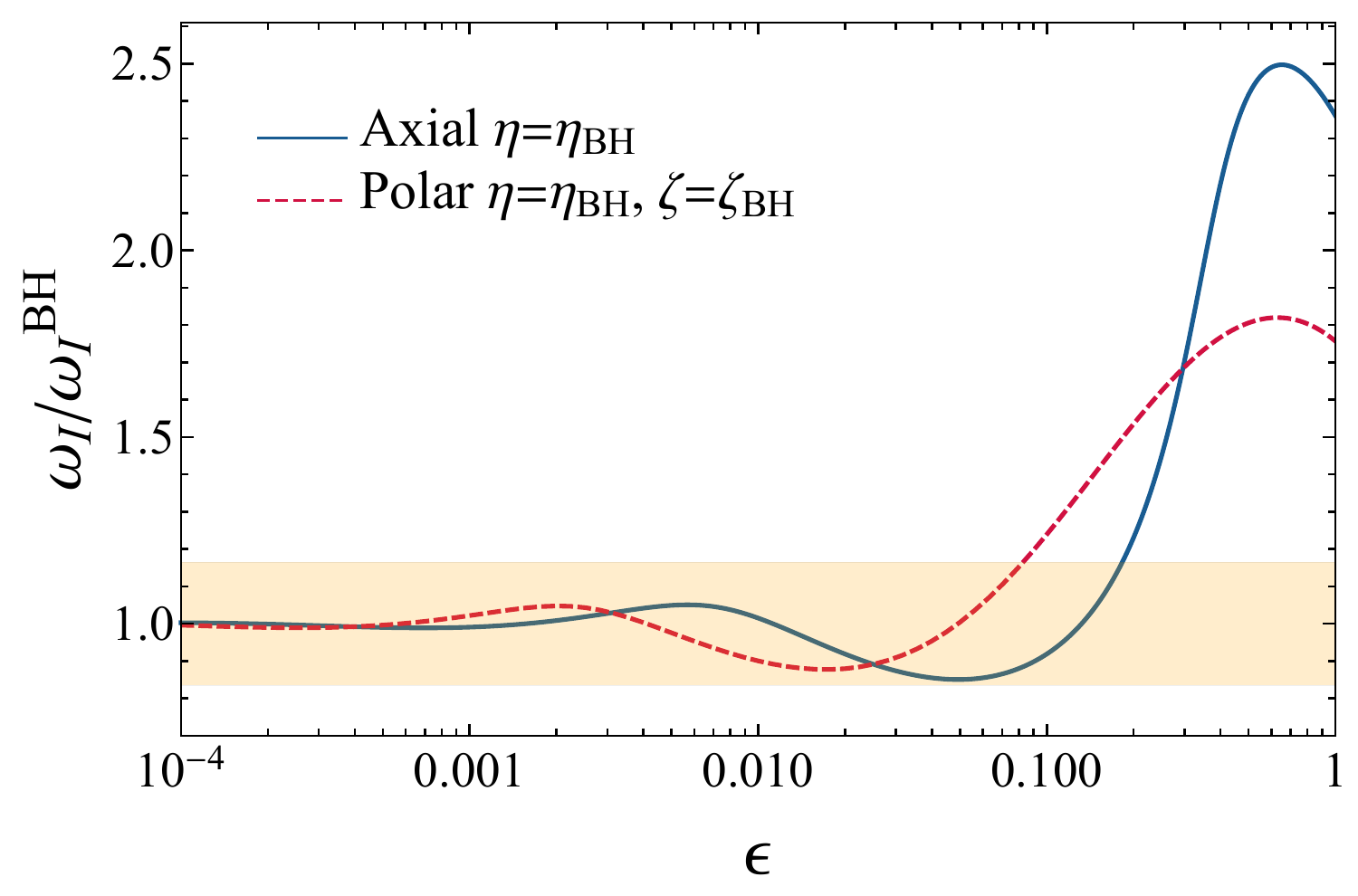}
\caption{Real (left panel) and imaginary (right panel) part of the QNMs of an ECO with the same reflective properties of a BH (described by a fictitious fluid with shear viscosity $\eta=\eta_{\rm BH}$ and bulk viscosity $\zeta=\zeta_{\rm BH}$) compared to the fundamental $l=2$ gravitational QNM of a Schwarzschild BH, as a function of the ECO radius, where $r_0 = 2M(1+\epsilon$)~\cite{Maggio:2020jml}. The highlighted region corresponds to the maximum deviation (with $90\%$ credibility) for the least-damped QNM in the event GW150914~\cite{TheLIGOScientific:2016src} with respect to the Kerr BH case~\cite{Ghosh:2021mrv}.}
\label{fig:ECOQNMs}
\end{figure}

Fig.~\ref{fig:ECOQNMs} shows the ratio between the QNM frequencies of an ECO with the same reflective properties of a BH ($\eta=\eta_{\rm BH}$, $\zeta=\zeta_{\rm BH} \equiv -1/(16\pi)$) and the fundamental $\ell=2$ QNM of a Schwarzschild BH as a function of the ECO compactness. As $\epsilon$ increases, the ECO QNMs start deviating from the BH QNM. The highlighted regions in Fig.~\ref{fig:ECOQNMs} correspond to the maximum allowed deviation (with $90\%$ credibility) for the least-damped QNM in the event GW150914~\cite{TheLIGOScientific:2016src} with respect to the Kerr BH case, and corresponds to $\sim  16\%$ and $\sim 33\%$ for the real and the imaginary part of the QNM, respectively~\cite{Ghosh:2021mrv}. Horizonless compact objects with $\epsilon \lesssim 0.1$ are compatible with current measurements. Future ringdown detections will allow us to set more stringent constraints on the nature of compact objects.
%

\subsection{Ergoregion instability of ECOs}
In a stationary spacetime, the ergoregion is the region where the Killing vector field $\partial_t=(1,0,0,0)$ becomes spacelike. The outer boundary of the ergoregion (sometimes called the ergosphere) coincides with the horizon for a Schwarzschild BH, whereas for a Kerr BH the ergoregion extends outside the horizon. In the ergoregion no static observers can exist, and  negative-energy state are allowed. This is the key mechanism of the
Penrose's process which allows to extract energy and angular momentum from a Kerr BH~\cite{Penrose:1969}. Let us consider a particle with energy at infinity $E_0$ decaying in two particles inside the ergoregion. One of the two particles can have a negative energy $E_1 < 0$ and therefore the other one must have energy larger than the initial value, $E_2 > E_0$. The event horizon forces the negative-energy particle to fall into the BH, while the positive-energy particle can escape at infinity and extract energy from the BH~\cite{Brito:2015oca}.

Spinning compact objects with an ergoregion but without an event horizon are prone to the so-called ergoregion instability. Assuming the horizonless object is non-dissipative, the negative-energy particle remains in orbital motion inside the ergoregion, since it cannot be absorbed. It is therefore energetically favorable to cascade towards more negative-energy states leading to a runaway instability. This infinite cascade can be prevented only if the compact object can efficiently absorb the negative-energy states.

The ergoregion instability was proved by Friedmann in ultracompact stars under scalar and electromagnetic perturbations~\cite{1978CMaPh..63..243F}
and analysed in uniform-density stars~\cite{CominsSchutz,1996MNRAS.282..580Y,Kokkotas:2002sf,Brito:2015oca,Moschidis:2016zjy}, gravastars~\cite{Chirenti:2008pf}, boson stars~\cite{Cardoso:2007az}, superspinars~\cite{Pani:2010jz} and ultracompact Kerr-like ECOs~\cite{Cardoso:2008kj,Maggio:2017ivp,Maggio:2018ivz}.
The origin of the ergoregion instability in horizonless ultracompact objects is due to the existence of long-lived modes. As shown in Fig.~\ref{fig:lowfrequencyQNMs}, the imaginary part of the QNM frequencies of a static ECO tends to zero in the limit of large compactness ($\epsilon \to 0$). In the rotating case, these modes can turn unstable due to the Zeeman splitting of the frequencies as function of the azimuthal number. In the small-spin limit, the QNM frequencies can be written as~\cite{Pani:2012bp}
\begin{equation} \label{zeeman}
    \omega_{R,I} = \omega_{R,I}^{(0)} + m \chi \omega_{R,I}^{(1)} + \mathcal{O}(\chi^2) \,,
\end{equation}
where $\omega_{R,I}^{(0)}$ are the real and the imaginary parts of the QNM frequencies in the static case and $\omega_{R,I}^{(1)}$ are the first order corrections to the QNM frequencies in the spin. For an ultracompact ECO with $\epsilon \ll 1$, $\omega_{I}^{(0)} \sim 0$ and the first order correction in Eq.~\eqref{zeeman} can turn the sign of the imaginary part of the frequency to be positive for a certain value of the azimuthal number. The symmetries $m \to -m$, $\omega \to - \omega^*$ guarantee that the ergoregion instability generically affects ECOs above a critical value of the spin.

\begin{figure}[t]
\centering
\includegraphics[width=0.49\textwidth]{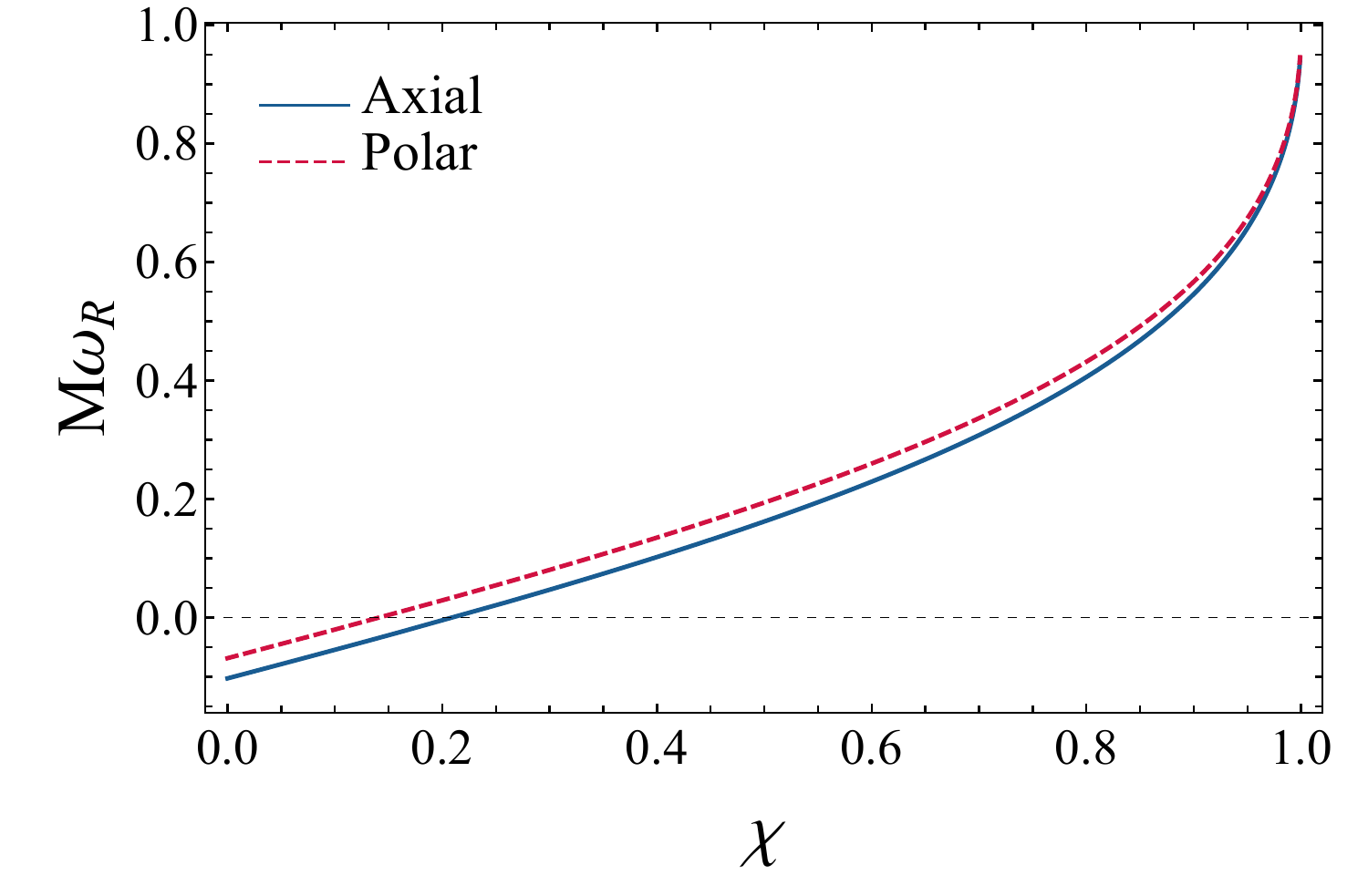}
\includegraphics[width=0.49\textwidth]{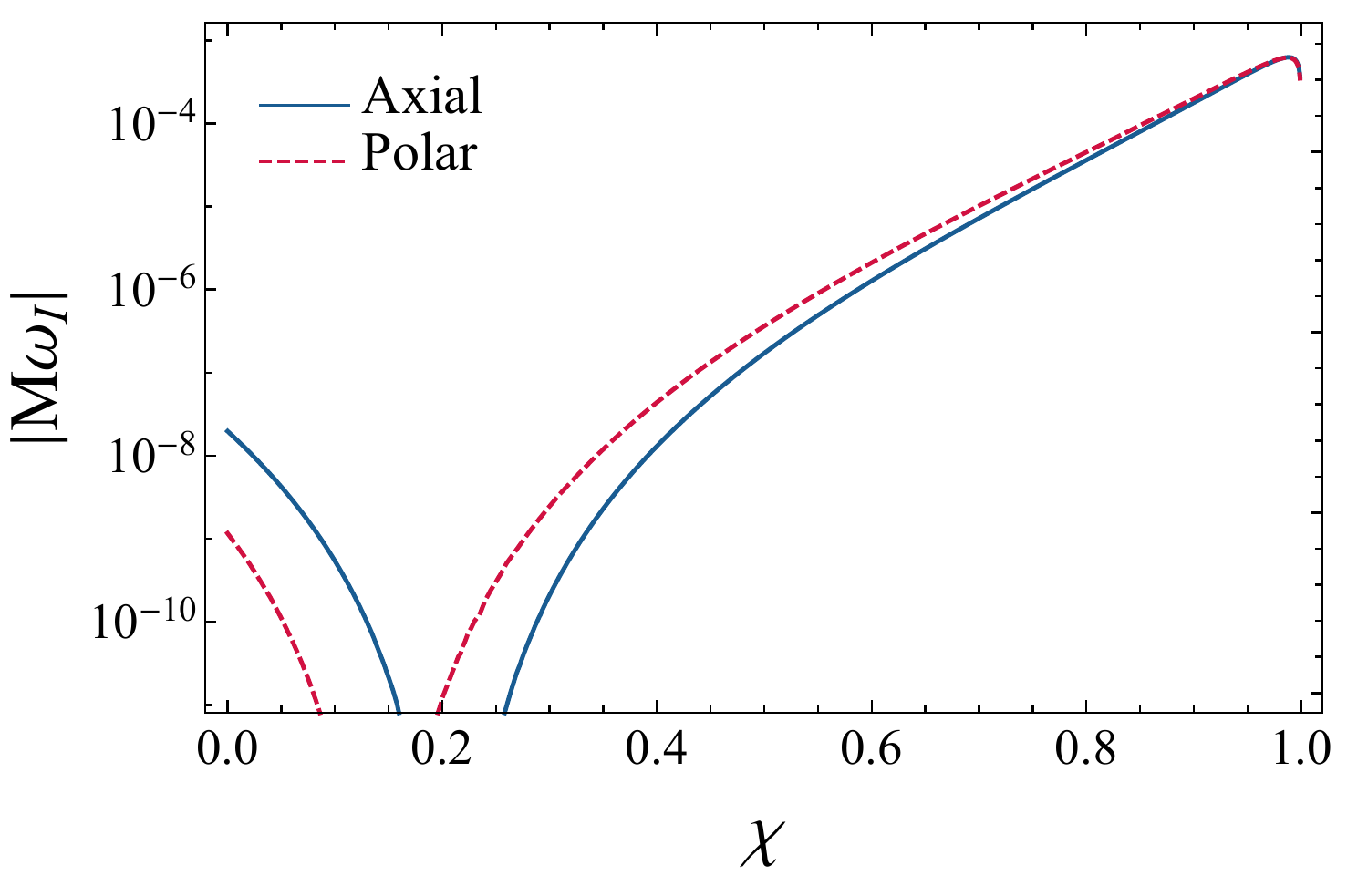}
\caption{Real (left panel) and imaginary (right panel) part of the fundamental gravitational ($\ell=m=2$) QNM of a Kerr-like ECO as a function of the spin. The surface of the ECO is located at $r_0 = r_+ \left(1 + \epsilon \right)$, where $\epsilon=10^{-10}$, and is perfectly reflecting,  $|\mathcal{R}|^2=1$. The ECO is affected by an ergoregion instability above a critical value of the spin that differs for axial and polar perturbations~\cite{Maggio:2018ivz}.} 
\label{fig:ergoregioninstability}
\end{figure}

Figure~\ref{fig:ergoregioninstability} shows the fundamental gravitational ($\ell=m=2$) QNM frequencies of a perfectly reflecting Kerr-like ECO as function of the spin with a given radius $r_0=r_+(1+\epsilon)$, where $r_+= M(1+\sqrt{1 - \chi^2})$ and $\epsilon=10^{-10}$~\cite{Maggio:2018ivz}. The real part of the QNM frequency has a zero crossing at some critical value of the spin which depends on the axial or polar nature of the perturbation. Most importantly, the imaginary part of the QNM frequency change sign above the same critical values of the spin, turning the ECO from stable into unstable.

For perfectly reflecting Kerr-like ECOs with $\epsilon \ll 1$, the critical value of the spin can be computed analytically from the generalization of Eqs.~\eqref{MwR},~\eqref{MwI} to the spinning case which are accurate when $\omega_R \simeq \omega_I \simeq 0$ (for details see Refs.~\cite{Maggio:2018ivz,Cardoso:2019rvt}). The ergoregion instability occurs for~\cite{Maggio:2018ivz, Cardoso:2019rvt}
\begin{equation}
    \chi>\chi_{\rm crit} \sim \frac{\pi}{m |\log \epsilon|} \left[ q+\frac{s(s+1)}{2} \right] \,,
\end{equation}
where $q$ is a positive even (odd) integer for axial (polar) modes. For example for an ECO with Planckian corrections at the horizon scale ($\epsilon=10^{-40}$), $\chi_{\rm crit} \simeq 0.03, 0.05$ for gravitational $\ell=m=2$ axial and polar perturbations, respectively. We conclude that even slowly spinning ECOs are unstable due to the ergoregion instability.

The timescale of the instability is defined as $\tau_{\rm inst} \equiv 1/|\omega_I|$. From Fig.~\ref{fig:ergoregioninstability}, for an ultracompact ECO with $\epsilon=10^{-10}$ and spin $\chi=0.7$, the instability timescale of the $\ell=m=2$ mode is
\begin{equation}
    \tau_{\rm inst} \in (5,7) \left( \frac{M}{10 M_\odot}\right) \ \rm s \,,
\end{equation}
where the lower (upper) bound is for polar (axial) perturbations. Let us notice that the low-frequency approximation of $\omega_I$ is not accurate for $\chi=0.7$, since $M\omega_R \sim 0.3$.
The ergoregion instability acts on a timescale which is short compared to the accretion timescale of astrophysical BHs, i.e., $\tau_{\rm Salpeter} \sim 4 \times 10^7 \ \rm yr$. However the instability timescale is longer than the decay time of the ringdown of BHs, i.e., $\tau_{\rm ringdown} \sim 0.5 \ \rm ms$ for a $10M_{\odot}$ object, and in general it is parametrically longer than the light-crossing time of the object. If the remnant of a compact binary coalescence was an ECO, the ergoregion instability could spin down the remnant over a timescale $\tau_{\rm inst}$ until the condition for the stability, $\chi=\chi_{\rm crit}$, is satisfied~\cite{Brito:2015oca}. This process would lead to a stochastic GW background due to spin loss~\cite{Fan:2017cfw,Du:2018cmp}. The absence of such background in the first observing run of Advanced LIGO already puts strong constraints on perfectly-reflecting ECOs which can be a small percentage of the astrophysical population~\cite{Barausse:2018vdb}.

\begin{figure}[t]
\centering
\includegraphics[width=0.65\textwidth]{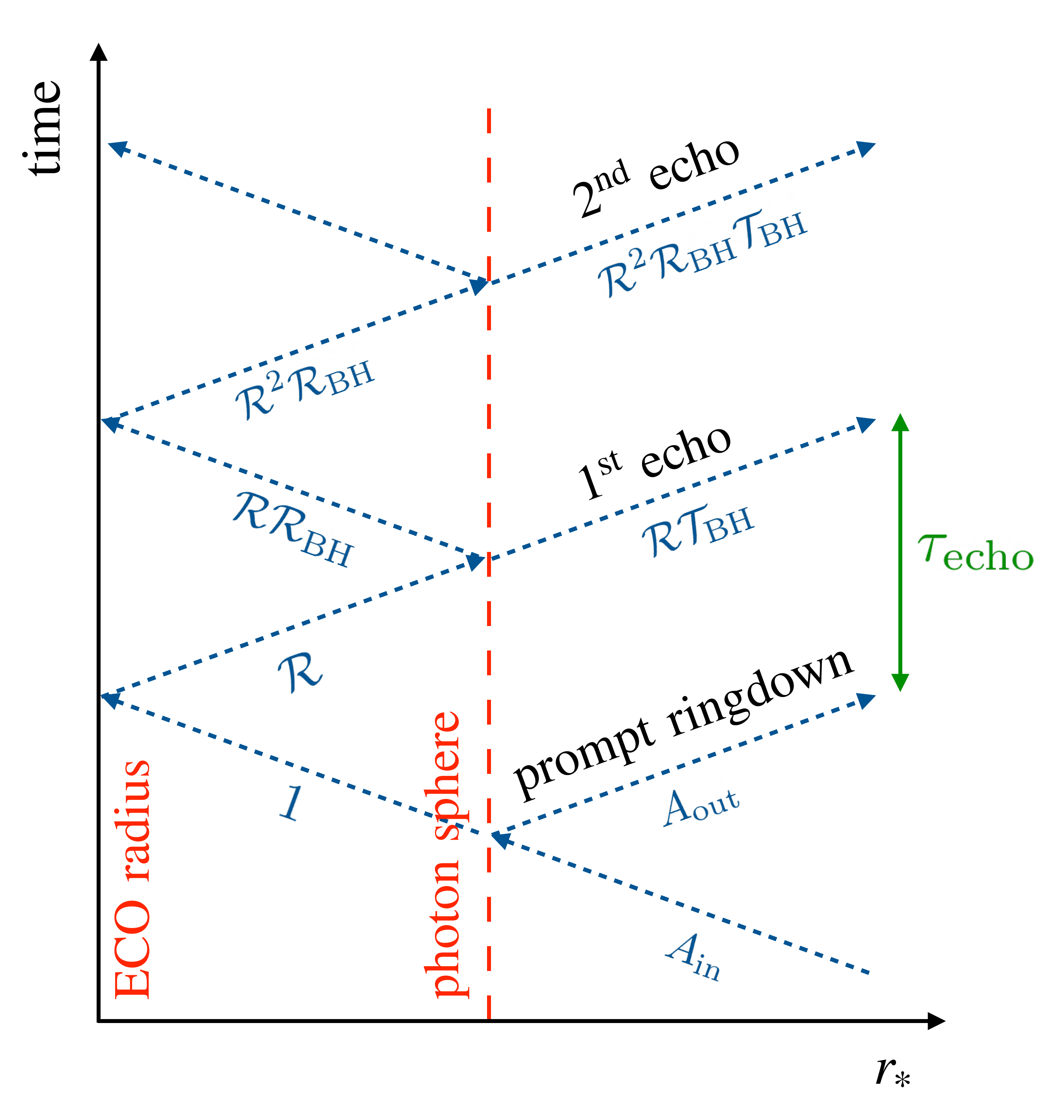}
\caption{Schematic diagram for the wave propagation in the spacetime of an ECO \cite{Vilenkin:1978uc,Abedi:2016hgu,Cardoso:2019rvt}.} 
\label{fig:cavityechoes}
\end{figure}

\begin{figure}[t]
\centering
\includegraphics[width=0.65\textwidth]{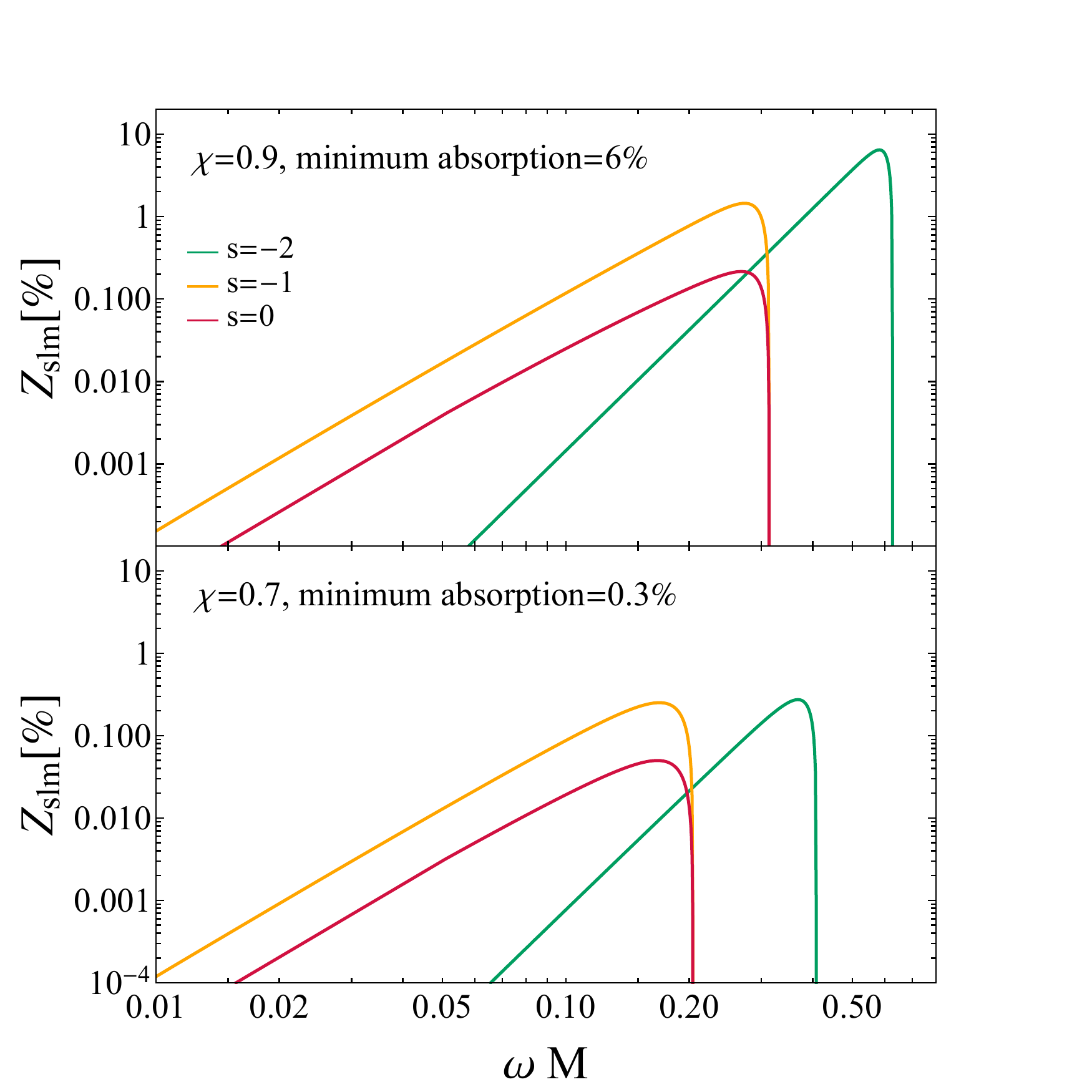}
\caption{Superradiant amplification factor of a BH as a function of the frequency for ($\ell=m=1$) scalar, electromagnetic and ($\ell=m=2$) gravitational perturbations. The minimum absorption rate to have a stable ECO for any type of perturbation is $0.3\%$ ($6\%$) for $\chi=0.7$ ($\chi=0.9$)~\cite{Maggio:2018ivz}.} 
\label{fig:Z}
\end{figure}

One way of quenching the ergoregion instability is by assuming that the surface of the ECO is partially absorbing~\cite{Maggio:2017ivp}. This model is more realistic than a perfectly reflecting surface since the compact object can absorb part of the radiation through viscosity, dissipation, fluid mode excitation, etc. The minimum absorption rate to have a stable ECO is related to the maximum amplification factor of BHs~\cite{Maggio:2018ivz}. In order to derive this result, let us analyse a spin-$s$ perturbation in the background of a spinning ECO. It is convenient to introduce the Detweiler's function which is governed by the master equation~\cite{1977RSPSA.352..381D}
\begin{equation} \label{eqDet}
    \frac{d^2 \Psi(r)}{dr_*^2} - V(r,\omega) \Psi(r) = 0 \,,
\end{equation}
where the effective potential reads
\begin{equation} \label{potentialDet}
    V(r,\omega) = \frac{U \Delta}{(r^2 + a^2)^2} + G^2 + \frac{dG}{dr_*} \,, \\
\end{equation}
with
\begin{eqnarray}
    G &=& \frac{s(r-M)}{r^2+a^2} + \frac{r \Delta}{(r^2+a^2)^2} \,, \\
    U &=& V_S + \frac{2 \alpha' + (\beta' \Delta^{s+1})'}{\beta \Delta^s} \,, \\
    V_S &=& -\frac{1}{\Delta} \left[ K^2 - i s \Delta' K + \Delta (2 i s K' - \lambda_s)\right]
\end{eqnarray}
where $\Delta=r^2-2Mr+a^2$, $K=(r^2+a^2)\omega-am$, the prime denotes a derivative with respect to $r$, $\alpha$ and $\beta$ are chosen such that the potential in Eq.~\eqref{potentialDet} is real~\cite{1977RSPSA.352..381D,Maggio:2018ivz} and $a=\chi M$. The two independent solutions of Eq.~\eqref{eqDet} have asymptotic behavior
\begin{equation} \label{asymptoticsplus}
\tilde \Psi_+(\omega, r_*) \sim 
\begin{cases}
 \displaystyle 
e^{+i \omega r_*} & \text{ as } r_* \to + \infty\\ 
 \displaystyle  
 B_{\rm out}(\omega)e^{+i k r_*}  +  B_{\rm in}(\omega) e^{- i k r_*} & \text{ as } r_* \to 
- \infty
\end{cases} \,,\\
\end{equation}
\begin{equation} \label{asymptoticsminus}
\tilde \Psi_-(\omega, r_*) \sim 
\begin{cases}
 \displaystyle 
 A_{\rm out}(\omega)e^{+i \omega r_*}  +  A_{\rm in}(\omega) e^{-i \omega r_*} & \text{ as } r_* \to + \infty \\ 
 \displaystyle  
 e^{-i k r_*} & \text{ as } r_* \to - \infty \\
\end{cases} \,,
\end{equation}
where the Wronskian of the solutions is conserved $W_{\rm BH}=\frac{d\tilde \Psi_+}{dr_*}\tilde \Psi_- -\tilde \Psi_+ \frac{d\tilde \Psi_-}{dr_*}=2 i k B_{\rm out}$,  $k=\omega-m \Omega$, and $\Omega=\chi/(2r_+)$ is the angular velocity of a Kerr BH at the event horizon. Let us define the reflection and trasmission coefficients of a wave coming from the left of the photon-sphere barrier with unitary amplitude as
\begin{equation} \label{RBH_TBH}
    \mathcal{R}_{\text{BH}} = \frac{B_{\rm in}}{B_{\rm out}} \,, \qquad \mathcal{T}_{\text{BH}} = \frac{1}{B_{\rm out}} \,.
\end{equation}
As shown in Fig.~\ref{fig:cavityechoes}, after each bounce in the cavity between the ECO surface and the photon sphere the perturbation acquires a factor $\mathcal{R}\mathcal{R}_{\rm BH}$, where $\mathcal{R}$ is the ECO surface reflectivity and $\mathcal{R}_{\rm BH}$ is defined in Eq.~\eqref{RBH_TBH}. Due to the conservation of the Wronskian, $|\mathcal{R}_{\rm BH}| = |A_{\rm out}/A_{\rm in}|$ where $A_{\rm in}$ and $A_{\rm out}$ are the coefficients of the incident and reflected wave, respectively, at the photon sphere for a left-moving wave originating at infinity.
The latter coefficients are related to the amplification factor of BHs for a wave of spin $s$ by
\begin{equation}
    Z_{slm} = \left| \frac{A_{\rm out}}{A_{\rm in}} \right|^2 - 1 \,.
\end{equation}
The condition for the energy in the cavity to grow indefinitely is $|\mathcal{R} \mathcal{R}_{BH}|^2>1$ which implies that the object is unstable due to the ergoregion instability if
\begin{equation} \label{ergoregionsuperradiance}
    |\mathcal{R}|^2 > \frac{1}{1+Z_{slm}} \,.
\end{equation}
Since the surface reflectivity is defined to be $|\mathcal{R}|^2 \leq 1$, Eq.~\eqref{ergoregionsuperradiance} implies that the ergoregion instability occurs when the real part of the QNM frequency is in the superradiant regime, i.e.,  $Z_{slm}>0$. In order to quench the ergoregion instability at any frequencies, the surface absorption, $1-|\mathcal{R}|^2$, needs to be larger than the maximum amplification factor of superradiance, namely
\begin{equation}
    1-|\mathcal{R}|^2 \gtrsim Z_{\rm max} \,,
\end{equation}
where $Z_{\rm max} \ll 1$. Figure~\ref{fig:Z} shows the amplification factor of a BH as a function of the frequency under scalar, electromagnetic and gravitational perturbations and for several values of the BH spin. In order to have a stable spinning ECO under any type of perturbation, the surface absorption needs to be at least $0.3\%$ ($6\%$) for an ECO with $\chi=0.7$ ($\chi=0.9$). Let us notice that the maximum amplification factor of an extremal BH is $\approx 138 \%$ for $\ell=m=2$ gravitational perturbations~\cite{Brito:2015oca,1974ApJ...193..443T} therefore an absorption rate of $\approx 60 \%$ would allow for stable ECOs with any spin~\cite{Maggio:2018ivz}. Some models of quantum BHs have a frequency-dependent reflectivity $\mathcal{R}=e^{-|k|/T_{\rm H}}$, where $T_{\rm H}$ is the Hawking temperature, which allows for stable spinning solutions against the ergoregion instability~\cite{Oshita:2019sat}.

\subsection{Gravitational-wave echoes}

GW echoes are an additional signal that would be emitted in the postmerger phase of a compact binary coalescence when the remnant is a horizonless ultracompact object~\cite{Kokkotas:1999bd,Kokkotas:1995av,Ferrari:2000sr,Cardoso:2016oxy} (see also~\cite{Tominaga:1999iy,Andrade:1999mj,Tominaga:2000cs,Andrade:2001hk} for related studies). Possible sources of GW echoes are near-horizon quantum structures~\cite{Cardoso:2016rao,Cardoso:2016oxy,Wang:2019rcf}, ultracompact neutron stars~\cite{Ferrari:2000sr,Pani:2018flj} and BHs in modified theories of gravity in which the graviton reflects effectively on a hard wall~\cite{Zhang:2017jze,Oshita:2018fqu}. The key features of the sources of GW echoes are the existence of a photon sphere in the exterior spacetime and the absence of an event horizon~\cite{Cardoso:2017cqb,Cardoso:2019rvt}. As shown in Fig.~\ref{fig:effectivepotential},
the effective potential of a perturbed ultracompact object displays a maximum near the photon sphere. If the radius of the object is smaller than the photon sphere, the effective potential features a cavity. If sufficiently compact, the latter can support quasi-trapped modes that leak out of the potential barrier through tunneling effects and are responsible for the GW echoes~\cite{Cardoso:2016rao}.

In order to describe the dynamical emission of GW echoes, let us analyse the scattering of a Gaussian pulse by the compact object. When the pulse crosses the photon sphere and perturbs it, a prompt ringdown signal is emitted at infinity as shown in Fig.~\ref{fig:cavityechoes}~\cite{Vilenkin:1978uc,Abedi:2016hgu,Cardoso:2019rvt}. The prompt ringdown emitted by an ultracompact horizonless object is almost indistinguishable from the BH ringdown because the photon sphere is approximately at the same location and has a similar shape~\cite{Cardoso:2016rao}. Afterwards, the perturbation travels inside the photon-sphere barrier and is reflected by the surface of the compact object. A fraction of the radiation is absorbed by the compact object depending on its reflective properties~\cite{Maggio:2018ivz,Burgess:2018pmm,Oshita:2019sat}. After each interaction with the photon sphere, a GW echo is emitted at infinity with a progressively smaller amplitude. 
The photon-sphere barrier acts as a frequency dependent high-pass filter. In particular the characteristic frequencies governing the prompt ringdown are very similar to the BH QNM frequencies, even if the latter are not part of the spectrum of horizonless ultracompact objects. The frequencies governing each subsequent GW echo become progressively smaller and at late times the GW signal is dominated by the low-frequency QNMs of the ECO~\cite{Mark:2017dnq,Wang:2018gin,Maggio:2019zyv}.

The morphology of the GW echoes gives us information about the properties of the ECO, in particular its compactness and its reflectivity. The delay time between subsequent GW echoes is associated with the light crossing time and depends logarithmically on the compactness of the ECO~\cite{Cardoso:2016rao,Cardoso:2016oxy}. For a nonspinning object (see Ref.~\cite{Abedi:2016hgu} for the spinning case),
\begin{equation}
    \tau_{\mathrm{echo}} = 2 M \left(1 - 2 \epsilon -2 \log \epsilon\right) \,,
\end{equation}
therefore the more the object is compact ($\epsilon \ll 1$), the (logaritmically) longer is the time delay between GW echoes. In principle, the logarithmic dependence on $\epsilon$ would allow to detect even Planckian corrections ($\epsilon \sim \ell_{\rm Planck}/M$) at the horizon scale few $\text{ms}$ after the merger for a remnant with $M\sim10M_\odot$. 

\begin{figure}[t]
\centering
\includegraphics[width=0.49\textwidth]{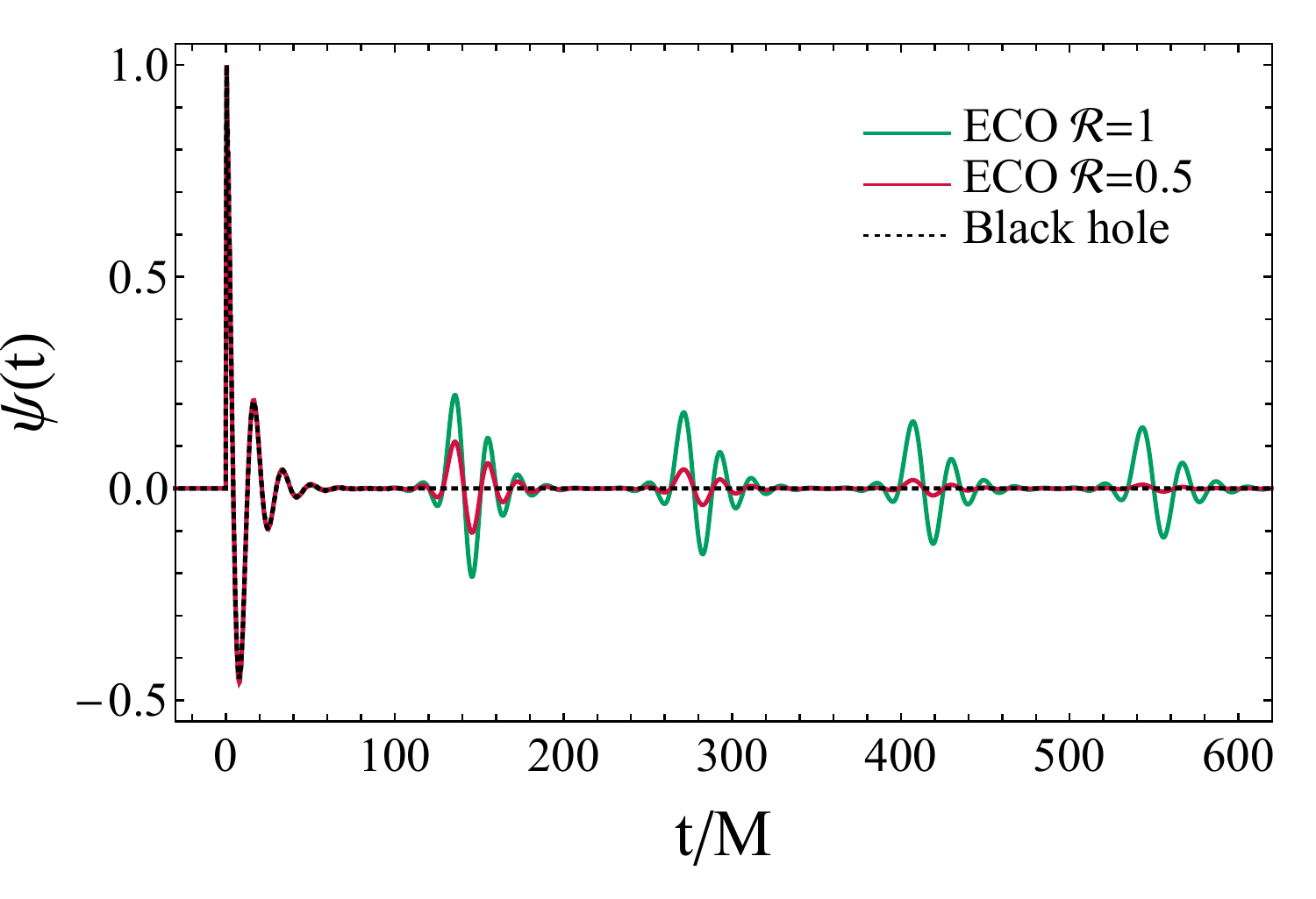}
\includegraphics[width=0.49\textwidth]{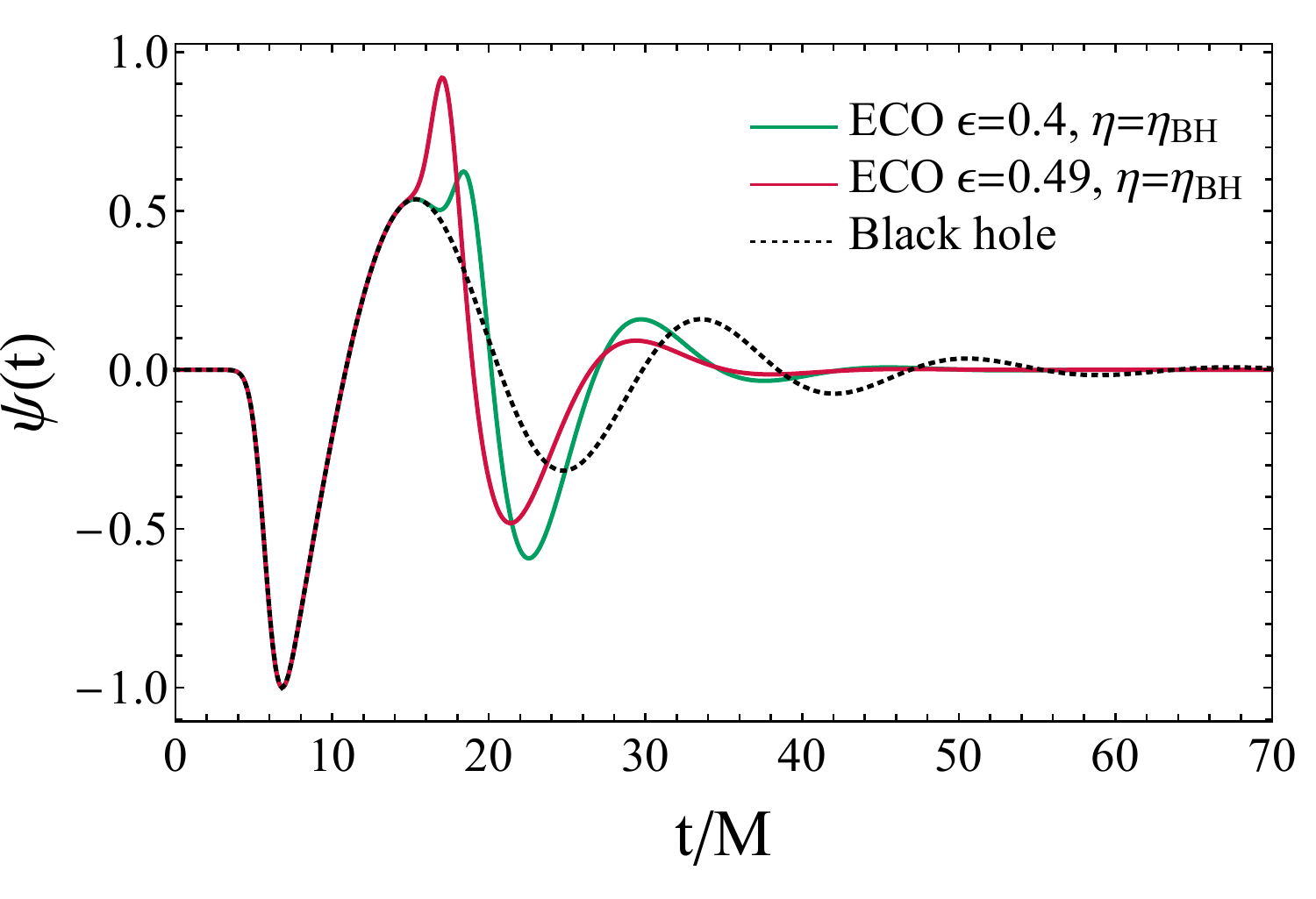}
\caption{Left panel: GW echoes emitted by a static ultracompact horizonless object with a given compactness ($\epsilon\simeq 5 \times 10^{-16}$) and several values of the surface reflectivity ($\mathcal{R}=1, 0.5$). Right panel: Ringdown of an ECO with radius $r_0 = 2M (1+\epsilon)$ and the same reflective properties of a BH~\cite{Maggio:2019zyv,Maggio:2020jml}.
} 
\label{fig:echoes}
\end{figure}

However, a crucial parameter that regulates the morphology of the echo signal (in particular the damping factor between subsequent echoes) is the reflectivity, as shown in the left panel of Fig.~\ref{fig:echoes}~\cite{Maggio:2020jml}. For a perfectly reflecting ECO the relative amplitude of the GW echoes is maximum, whereas it decreases for a  partially absorbing ECO and it goes to zero in the limit of a perfectly absorbing object as in the BH case.

When the remnant of a merger is an ECO with $\epsilon \gtrsim 0.01$, the delay time of the GW echoes is comparable with the decay time of the prompt ringdown where the latter is associated to the decay time of the fundamental QNM of a BH, $\tau_{\rm damping} \approx 10 M$. 
As a consequence, the prompt ringdown emitted by the direct excitation of the photon sphere interferes with the first GW echo in an involved pattern as shown in the right panel of Fig.~\ref{fig:echoes}~\cite{Maggio:2020jml}. In particular when the two pulses sum in phase, the interference produces high peaks in the GW signal. Furthermore, subsequent echoes are suppressed because the cavity between the photon sphere and the radius of the compact object is small and does not trap the modes efficiently.

Let us derive the gravitational waveform that would be emitted by an ultracompact horizonless object in the ringdown. We require the GW signal emitted at infinity to be a purely outgoing wave, $\tilde{\psi}(\omega, r_* \rightarrow \infty) = \tilde{Z}^+(\omega) e^{i \omega r_*}$. In frequency domain the GW signal emitted by an ECO can be written in terms of the GW signal that would be emitted by a BH and reprocessed by a transfer function. In particular~\cite{Mark:2017dnq}
\begin{equation} \label{ringdownECO}
    \tilde{Z}^+(\omega) = \tilde{Z}^+_{\text{BH}}(\omega) + \mathcal{K}(\omega) \tilde{Z}^-_{\text{BH}}(\omega) \,,
\end{equation}
where $\tilde{Z}^{\pm}_{\text{BH}}(\omega)$ are the responses of a BH at infinity and near the horizon, for the plus and minus sign respectively, to a source $\tilde{\mathcal{S}}$
\begin{equation}
    \tilde{Z}^{\pm}_{\text{BH}}(\omega) = \frac{1}{W_{\text{BH}}} \int_{-\infty}^{+\infty} dr_* \tilde{\mathcal{S}} \Psi_{\mp} \,,
\end{equation}
where $\Psi_{\pm}$ are the independent solutions of the homogeneous equation~\eqref{eqDet} with asymptotics in Eqs.~\eqref{asymptoticsplus},~\eqref{asymptoticsminus}. The transfer function is defined as~\cite{Mark:2017dnq,Cardoso:2019rvt}
\begin{equation} \label{transferfunction}
    \mathcal{K}(\omega) = \frac{\mathcal{T}_{\rm BH} \mathcal{R}(\omega) e^{-2 i k r_*^0}}{1 - \mathcal{R}_{\rm BH} \mathcal{R}(\omega) e^{-2 i k r_*^0}} .
\end{equation}
According to Eq.~\eqref{ringdownECO}, the GW signal emitted by an ECO is the same as the one emitted by a BH with an extra GW emission that depends on the reflectivity of the ECO. In order to get an insight of the additional GW emission, let us expand the transfer function in Eq.~\eqref{transferfunction} as a geometric series~\cite{Mark:2017dnq,Correia:2018apm}
\begin{equation} \label{transferfunctionseries}
    \mathcal{K}(\omega) = \mathcal{T}_{\rm BH} \mathcal{R}(\omega) e^{-2 i k r_*^0} \sum_{j=1}^{\infty} \left[ \mathcal{R_{\rm BH}} \mathcal{R}(\omega)\right]^{j-1} e^{-2 i (j-1) k r_*^0} \,.
\end{equation}
In view of Eq.~\eqref{transferfunctionseries}, the GW signal takes the form of a series of pulses where the index $j$ represents the signal emitted by the $j$-th echo. The phase factor $2 i k x_0$ corresponds to the time delay between each pulse due to the round-trip time between the photon sphere and the surface of the ECO.
Each echo can have a phase inversion relative to the previous one when the factor $\mathcal{R}_{\rm BH} \mathcal{R}(\omega)$ has a negative sign~\cite{Testa:2018bzd,Maggio:2019zyv}. Eq.~\eqref{ringdownECO} allows to construct a template for the GW signal emitted by an ECO that depends only on the black-hole ringdown parameters and the parameters of the ECO, i.e.,  its compactness and reflectivity. In time domain, the gravitational waveform is obtained by an inverse Fourier transform,
\begin{equation}
    h(t) = \frac{1}{\sqrt{2 \pi}} \int_{-\infty}^{+\infty} d\omega \tilde{Z}^+(\omega) e^{-i \omega t} \,.
\end{equation}

Several phenomenological templates have been developed in order to perform matched-filter searches of GW echoes~\cite{Cardoso:2019rvt,Abedi:2020ujo}. In time domain some templates are based on standard GR ringdown templates with extra parameters that are related to the morphology of the GW echoes, i.e., the delay time and the damping factor~\cite{Abedi:2016hgu,Nakano:2017fvh}. Several time-domain templates approximate the GW echoes by complex Gaussians~\cite{Wang:2018gin} and by a superposition of sine-Gaussians with free parameters~\cite{Maselli:2017tfq}.
In frequency domain some analytical templates depend explicitly on the physical parameters ECOs~\cite{Mark:2017dnq} and are obtained with analytical approximations of the transfer function in Eq.~\eqref{transferfunction}~\cite{Testa:2018bzd,Maggio:2019zyv}.

Some searches for GW echoes have been performed~\cite{Abedi:2020ujo}. A tentative evidence for GW echoes has been claimed in the postmerger phase of  compact binary coalescences detected by Advanced LIGO and Advanced Virgo in  the first two observing runs~\cite{Abedi:2016hgu,Conklin:2017lwb,Abedi:2018npz}. However the statistical significance of GW echoes has been claimed to be low and consistent with noise~\cite{Westerweck:2017hus,Nielsen:2018lkf}. Some independent searches confirmed to find no evidence for GW echoes based on morphology-independent searches with a decomposition of the signal in terms of generalized wavelets~\cite{Tsang:2018uie,Tsang:2019zra} and template-based searches~\cite{Uchikata:2019frs,Lo:2018sep}. Moreover, no evidence for GW echoes has been found in the binary BH events from the GWTC-2 catalog~\cite{Abbott:2020jks} confirming previous results.
Third-generation detectors like the Einstein Telescope~\cite{Hild:2010id,Maggiore:2019uih}, the Cosmic Explorer~\cite{Reitze:2019iox} and the Laser Interferometer Space Antenna~\cite{Audley:2017drz} will allow to detect GW echoes even for objects with small reflectivity, or to put strong constraints on ECO models, given the large signal-to-noise-ratio in the ringdown of $\mathcal{O}(100)$~\cite{Testa:2018bzd,Maggio:2019zyv}.

\section{Conclusions and Open Issues}
We conclude with a list of some of the most outstanding open problems in this area:

\begin{itemize}
    \item The formation channel of ECOs is mostly unmodeled, except for boson stars~\cite{Seidel:1993zk,Liebling:2012fv}. For ultracompact ECOs it relies on the idea that some quantum effects can prevent the formation of the event horizon, however there are no available simulations. 
    \item Beside the case of boson stars, very little is known about the dynamics of isolated and, especially, ECO binaries. Numerical simulations would be crucial to study the stability of ECOs, as well as ECO mergers and to develop consistent inspiral-merger-ringdown waveform templates.
    \item Many studies so far have assumed the ECO reflectivity to be constant, but in realistic models ${\cal R}$ is a complex function of the frequency and the spin. Modeling this property is important for both the viability and the phenomenology of realistic models~\cite{Burgess:2018pmm,Oshita:2019sat}.
    \item It would be interesting to extend the membrane paradigm for ECOs~\cite{Maggio:2020jml} to spinning objects and to other configurations.
    \item Phenomenological studies of fuzzballs are in their infancy and should be extended in various directions; e.g., the ringdown, tidal effects, and the impact of the multipole moments on the inspiral waveform are uncharted territories.
\end{itemize}

\section{\textit{Acknowledgments}}
We acknowledge financial support provided under the European Union's H2020 ERC, Starting 
Grant agreement no.~DarkGRA--757480. We also acknowledge support under the MIUR PRIN and FARE programmes (GW-NEXT, CUP:~B84I20000100001) and networking support by the COST Action CA16104 and 
support from the Amaldi Research Center funded by the MIUR program ``Dipartimento di 
Eccellenza" (CUP:~B81I18001170001).


\bibliographystyle{spbasic} 
\bibliography{reference}
\printindex
\end{document}